\documentclass[aps,prb,reprint,superscriptaddress,floatfix,longbibliography,nofootinbib]{revtex4-2}

\usepackage[colorlinks,citecolor=blue,urlcolor=blue,linkcolor=blue]{hyperref}
\usepackage{amsmath,amssymb,amsthm,graphicx,bbm,bm}
\usepackage{braket}
\usepackage{mathtools}
\usepackage{physics}
\usepackage{dsfont}
\usepackage{enumitem}
\usepackage{xcolor}
\usepackage{placeins}

\newcommand{\ii}{\mathrm{i}}

\makeatletter
\AtBeginDocument{\immediate\write\@auxout{\string\citation{apsrev42Control}}}
\makeatother

\begin{document}

\title{Spontaneous Symmetry Breaking in Chiral Current-Carrying High-Energy Eigenstates}

\author{Tigran A.~Sedrakyan}
\affiliation{Department of Physics, University of Massachusetts, Amherst, MA 01003, USA}
\affiliation{A.~Alikhanyan National Science Laboratory, Br.~Alikhanian 2, Yerevan 0036, Armenia}

\author{Jun-Jun Pang}
\affiliation{Department of Physics, University of Massachusetts, Amherst, MA 01003, USA}
\affiliation{Department of Physics, Nanjing University, Nanjing 210093, China}

\author{Chenan Wei}
\affiliation{A.~Alikhanyan National Science Laboratory, Br.~Alikhanian 2, Yerevan 0036, Armenia}
\affiliation{Department of Physics, University of Massachusetts, Amherst, MA 01003, USA}

\author{Baigeng Wang}
\affiliation{Department of Physics, Nanjing University, Nanjing 210093, China}

\begin{abstract}
Typical finite-energy-density eigenstates of nonintegrable systems are expected to obey the eigenstate thermalization hypothesis and to reproduce thermal local observables. In integrable systems, typical states are instead finite-entropy Bethe macrostates, or generalized Gibbs ensembles, with smooth quasiparticle occupations. Here we show that the spin-$\tfrac12$ XXX Heisenberg chain contains an exactly solvable exception. By biasing mutually commuting conserved charges, we use the ground state of a selector Hamiltonian to construct rare eigenstates of the undeformed XXX Hamiltonian. These states are atypical ordered, chiral, current-carrying, critical, zero-entropy Bethe macrostates at tunable XXX energy density. They lie outside the finite-entropy manifold dominating generalized Gibbs ensembles, yet remain exact XXX eigenstates with sharp Bethe occupations, finite scalar chirality, nonthermal local observables, and gapless Luttinger-liquid correlations. A finite-interval thermodynamic Bethe ansatz reveals an asymmetric chiral Bethe sea whose zero-field limit selects an extensive spin sector through $SU(2)$ symmetry breaking. A commuting exchange bias moves the same ordered macrostate through the full thermodynamic XXX energy band, including the Hilbert-space trace center. Finite-size DMRG and exact-diagonalization benchmarks indicate that the magnetochiral signatures persist under weak integrability breaking over prethermal time scales. The construction provides a controlled integrable realization of ETH-violating, scar-like large-deviation eigenstates and a route to ordered critical matter deep inside a many-body spectrum.
\end{abstract}
\maketitle

\section{Introduction}

A standard expectation for many-body spectra is that a generic finite-energy-density eigenstate of a nonintegrable Hamiltonian reproduces, for local observables, the predictions of the microcanonical or canonical ensemble at the same energy density.  This is the content of the eigenstate thermalization hypothesis (ETH), which explains how isolated nonintegrable systems can exhibit local equilibration without coupling to an external bath~\cite{deutsch1991,srednicki1994,rigol2008,dalessio2016,PatilRigol2025PRB}. Integrable systems do not generally thermalize to an ordinary Gibbs ensemble, because their extensive set of conserved charges imposes additional constraints. Their typical stationary states are instead generalized Gibbs ensembles (GGEs), equivalently Bethe macrostates described by smooth root or occupation functions and by a nonzero Yang--Yang entropy density~\cite{rigol2007,CassidyClarkRigol2011,ilievski2016,vidmar2016,Yuzbashyan2016,Reiter2021}. In both cases, the states that dominate the bulk of the many-body spectrum are finite-entropy macrostates. They are therefore not expected to resemble ordered zero-entropy ground states: uniform magnetization, parity-odd chirality, and coherent spin correlations should either vanish by symmetry or be strongly suppressed near the center of the band, except in specially constructed atypical eigenstates.

The most interesting physics often sits precisely at the boundary of this expectation.  Quantum many-body scars show that special high-energy states can evade ETH in clean systems~\cite{bernien2017,turner2018,serbyn2021,odea2025}.  Many-body localization and localization-protected order provide a disorder-based route, using emergent local integrals of motion~\cite{nandkishore2015,altman2015,abanin2019MBL}.  Far-from-equilibrium field theories exhibit nonthermal fixed points where collective structure survives outside ordinary equilibrium~\cite{Berges2008}.  Large-deviation physics supplies a further organizing principle: rare states can be exposed by tilting the statistical weight with respect to conserved currents or charges, as in current-biased ensembles and macroscopic fluctuation theory~\cite{BertiniMFT2015}.  These ideas suggest a sharp question for quantum integrable models: can one construct, solve, and physically interpret an atypical Bethe macrostate that is a genuine eigenstate, has zero entropy density, carries a persistent current, remains critical, and chooses a spin sector?

The spin $s=\tfrac12$ XXX chain provides a setting where this question can be answered exactly.  Its eigenstates are characterized not only by their energy, but by the simultaneous eigenvalues of an infinite family of mutually commuting conserved charges.  These charges, conventionally denoted $Q_n$, include the XXX Hamiltonian and higher local or quasilocal charges.  In the Bethe-ansatz description, the corresponding quantum numbers are fixed by the Bethe rapidities, or equivalently in the thermodynamic limit by Bethe root densities.  Conserved-charge data therefore give a much finer classification of eigenstates than the XXX energy density only by itself. 

We use this structure to construct and solve a rare sector of the spin $s=\tfrac12$ XXX spectrum.  The central result is that a conserved-charge selector gives an exact construction of atypical ordered, current carrying, critical, zero-entropy Bethe macrostates at arbitrary XXX energy density.  The selected state is an exact eigenstate of the original XXX Hamiltonian, but it is singled out by fixing additional conserved-charge eigenvalues rather than by minimizing the XXX energy.  It has finite energy density, vanishing Yang--Yang entropy density, finite scalar chirality and spin-sector order, nonthermal local observables, and a gapless Luttinger-liquid spectrum.  It is therefore an exactly solvable atypical Bethe macrostate lying outside the finite-entropy manifold that dominates generalized Gibbs ensembles.  In this sense, the construction identifies rare macrostates selected by higher conserved charges.  Their rarity is precisely what makes them a controlled integrable counterpart of ETH-violating and scar-like eigenstates: they are exceptional finite-energy-density eigenstates, but their exceptionality is determined directly from their conserved-charge quantum numbers and Bethe roots, rather than inferred from finite-size spectral anomalies.

The idea is a nonequilibrium-from-equilibrium construction.  In an integrable model, mutually commuting charges can be biased by static Lagrange multipliers.  The biased Hamiltonian has the same eigenvectors as the original Hamiltonian but ranks them differently. Its ground state therefore selects a particular high-energy eigenstate of the undeformed problem.  This is the zero-temperature, microcanonical end point of a large-deviation tilt in charge space.  It also clarifies the relation to GGEs and quantum scars.  A finite-temperature GGE describes a typical finite-entropy saddle compatible with specified conserved-charge densities. The selector instead takes the bias to zero generalized temperature and lands on an extremal, zero-entropy Bethe macrostate.  The construction is scar-like in the operational sense that it isolates exceptional nonthermal eigenstates, but it is not a conventional finite-dimensional scar tower inside a nonintegrable spectrum.  It is an exactly solvable Bethe large-deviation state.  The advantage is conceptual and practical: an excited-state problem for the original chain becomes an equilibrium Bethe-ansatz problem for an auxiliary Hamiltonian, while the selected state remains an eigenstate of the original dynamics.

For the $s=1/2$ XXX chain, we use the simplest commuting selector,
\begin{equation}
\label{eq:HJchih-intro}
  H_{J\chi h}=JQ_2+\alpha Q_3-hS^z .
\end{equation}
Here
\begin{equation}
\label{eq:HXXX}
H_{\rm XXX}(J)\equiv JQ_2,
\qquad
Q_2=\sum_j \mathbf S_j\!\cdot\!\mathbf S_{j+1},
\end{equation}
is the XXX spin-exchange Hamiltonian, $S^z=\sum_jS_j^z$ is the conserved total magnetization, and
\begin{equation}
\label{eq:Q3-intro}
Q_3=-\sum_j \mathbf S_j\!\cdot\!(\mathbf S_{j+1}\times\mathbf S_{j+2})
\end{equation}
is the scalar-chirality charge, equivalently the local energy-current charge of the XXX hierarchy up to normalization.  The coupling $\alpha$ selects a handed current sector.  The field $h$ is a probe used in the standard order of limits for spontaneous symmetry breaking (SSB): the thermodynamic limit is taken first, and only afterward $h\to0^\pm$.

\begin{figure}[t]
    \centering
    \includegraphics[width=\linewidth]{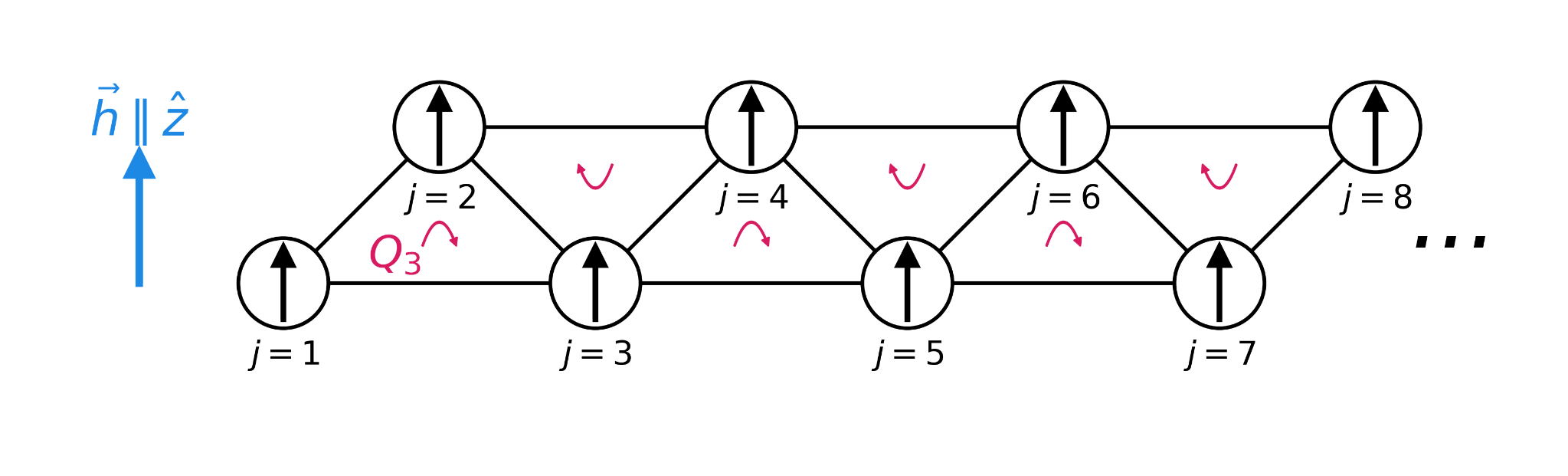}
    \caption{Schematic representation of the minimal current selector.  The field $h$ couples to the conserved spin generator $S^z$, while the three-spin charge $Q_3$ selects a scalar-chirality, or energy-current, sector.}
    \label{fig:lattice}
\end{figure}

The minimal current selector is obtained by setting $J=0$,
\begin{equation}
\label{eq:2}
  H_{\chi h}=\alpha Q_3-hS^z .
\end{equation}
We show it schematically in Figure~\ref{fig:lattice}.  This Hamiltonian is built from a conserved current operator of the same XXX integrable hierarchy.  Consequently, its ground state is a definite XXX eigenstate, selected by chirality and magnetization rather than by the ordinary exchange energy.

The finite-interval thermodynamic Bethe ansatz (TBA) developed in the present work reveals an asymmetric Bethe sea in the ground state of the minimal selector.  For the pure current selector at zero field we find the magnetization density and chirality density to be
\begin{eqnarray}
&&M(h=0^\pm)=\pm0.215636\ldots,
\nonumber\\
&&\chi(h=0)=-0.245150\ldots .
\end{eqnarray}
This is the central symmetry-breaking result in its simplest form.  The Hamiltonian $H_{\chi0}=\alpha Q_3$ has no explicit spin-axis field and is $SU(2)$ invariant on every finite chain.  Nevertheless, after the thermodynamic limit has been taken, an infinitesimal $h\to0^\pm$ selects a state with finite magnetization density.  The same state carries a macroscopic scalar chirality, so the selected ground-state sector of the minimal selector is spin-sector ordered, parity- and time-reversal odd, current carrying, and critical.  Because $Q_3$ commutes with $Q_2$, this ordered ground state of the selector is simultaneously a high-energy eigenstate of the undeformed XXX Hamiltonian.

\begin{figure*}[t]
    \centering
    \includegraphics[width=0.88\textwidth]{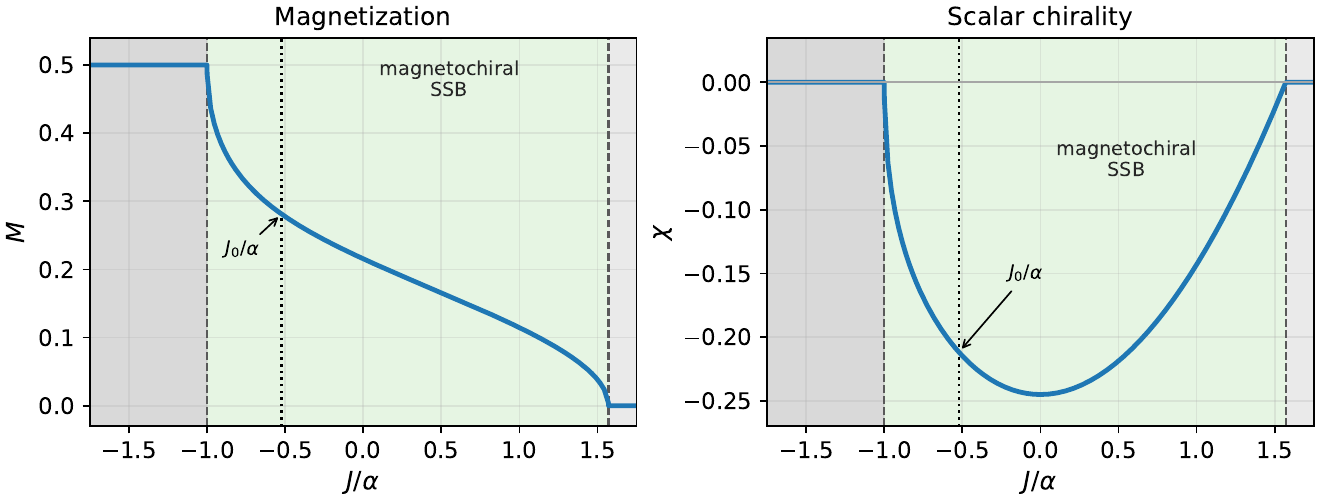}
    \caption{Order parameters of the zero-field selector $H_{J\chi0}=JQ_2+\alpha Q_3$ at $h=0^+$ and $\alpha=1$.  Left: magnetization density $M$.  Right: scalar-chirality density $\chi$.  The point $J=0$ is the minimal pure-current selector $H_{\chi0}=\alpha Q_3$.  The green shaded region, $-1<J/\alpha<\pi/2$, is the magnetochiral symmetry-broken branch with finite magnetization and finite scalar chirality.  The right gray region is the no-SSB XY-paramagnetic sector connected to the antiferromagnetic XXX sea, with no spontaneous spin polarization and no chirality at zero field.  The left gray region is the fully polarized ferromagnetic phase, which breaks spin rotation in the conventional way but has $\chi=0$.  The dotted line marks $J_0/\alpha=-0.52207095\ldots$, where the selected state has $e_{\rm XXX}=0$ while retaining the finite order parameters in Eq.~\eqref{eq:center-params}.}
    \label{fig:Jorders}
\end{figure*}

In Figure~\ref{fig:Jorders} we summarize the zero-field order parameters of the generalized selector $H_{J\chi0}=JQ_2+\alpha Q_3$.  The point $J=0$ is the minimal current selector defined above, so the finite value of $M$ at that point is the spontaneous magnetization just reported.  More importantly, the figure shows that this is not an isolated phenomenon.  The one-sided SSB occurs across the entire parameter range in which the selector ground state is chiral, namely throughout the finite-interval branch $-1<J/\alpha<\pi/2$.  At the lower boundary the chiral sea collapses into the ferromagnetic sector; at the upper boundary it merges with the antiferromagnetic XXX sea and loses zero-field chirality.  Inside the window, however, finite scalar chirality and spontaneous spin-sector selection coexist throughout the branch.

\begin{figure*}[t]
    \centering
    \includegraphics[width=0.86\textwidth]{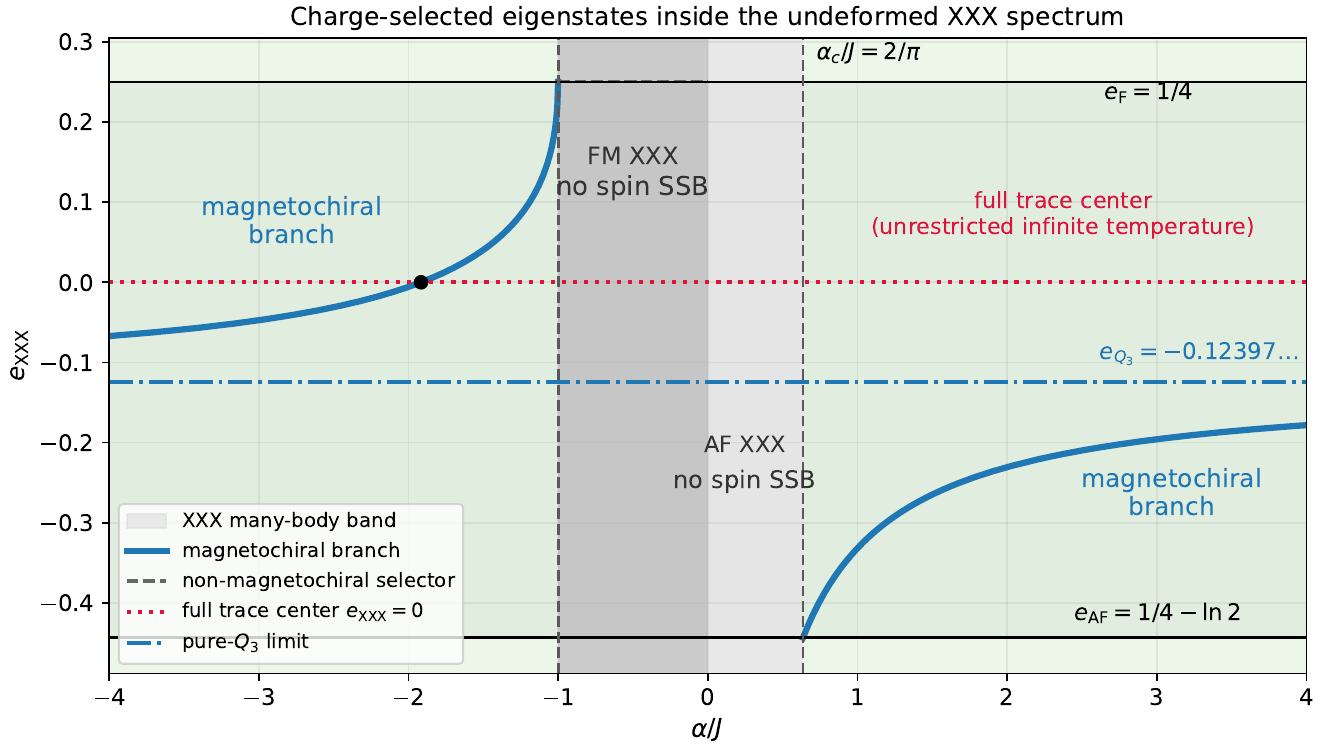}
    \caption{Charge-selected eigenstates inside the thermodynamic energy band of the undeformed XXX Hamiltonian $Q_2$, shown as a function of $\alpha/J$ at $h=0^+$.  The many-body band is bounded by $e_{\rm AF}=1/4-\ln2$ and $e_{\rm F}=1/4$.  The dotted red line is the full Hilbert-space trace center $e_{\rm XXX}=0$, i.e., the unrestricted infinite-temperature energy density.  The blue dash-dotted line is the pure-current limit $J\to0$, where $e_{Q_3}=-0.12396686\ldots$.  Green shading denotes the parameter region for the magnetochiral branch.  The two gray parameter regions represent parameter values where the ground state of the selector Hamiltonian is Heisenberg-like and thus is non-magnetochiral ferromagnetic (FM) XXX and antiferromagnetic (AF).  The black dot marks the zero-field trace-center tuning, where the selected eigenstate keeps finite magnetization and scalar chirality.}
    \label{fig:energy-alphaoverJ}
\end{figure*}

In terms of the undeformed XXX chain, the chiral parameter window in Figure~\ref{fig:Jorders} has a striking spectral meaning.  In Figure~\ref{fig:energy-alphaoverJ} we summarize the XXX energy density $e_{\rm XXX}$ of the selected state, measured with the undeformed exchange charge $Q_2$.  The gray band is the thermodynamic many-body energy band, bounded by the antiferromagnetic density $e_{\rm AF}=1/4-\ln2$ and the fully polarized density $e_{\rm F}=1/4$.  The dotted red line is the full Hilbert-space trace center $e_{\rm XXX}=0$, i.e., the unrestricted infinite-temperature energy density.  The blue curves are sharp charge-selected Bethe macrostates.  The chiral SSB window in Figure~\ref{fig:Jorders} maps onto the entire energy band shown in Figure~\ref{fig:energy-alphaoverJ}: the positive-$J$ branch reaches from the antiferromagnetic edge to the pure-$Q_3$ energy, while the negative-$J$ branch reaches from the pure-$Q_3$ energy to the ferromagnetic edge.  Tuning the commuting exchange bias can even place the ordered, current-carrying state at the trace center while preserving finite magnetization and chirality.

The same solution also explains why this high-energy state is nonthermal.  The selected Bethe occupations are exactly filled or empty, so the Yang--Yang entropy density vanishes point by point.  At the same time, particle-hole excitations at the Fermi edges form a $c=1$ Luttinger liquid.  Because the background carries chirality, the two Fermi velocities are generally unequal.  The state therefore resembles a zero-temperature critical fluid, but it sits at a high XXX energy density and carries a persistent handed current.  This links the present construction to broader phenomena in quantum physics: chiral Luttinger liquids at quantum Hall edges~\cite{Wen1990ChiralLL}, chiral spin liquids and scalar-chirality order in frustrated magnets~\cite{motrunich2006,greiter2014}, and ballistic or anomalous transport in one-dimensional quantum matter.

The construction is also experimentally relevant.  Exact integrability makes the analytic solution possible, but weak integrability breaking should produce a prethermal time window rather than immediately erasing the selected structure.  Our finite-size DMRG and exact-diagonalization benchmarks with a weak Ising perturbation show that the one-sided magnetization, the $c=1$ entanglement scaling, and the short-time fidelity remain visible on accessible system sizes.  Synthetic gauge fields, triangular or zigzag optical-lattice geometries, Rydberg arrays, and trapped-ion platforms provide natural ways to engineer or approximate the scalar-chirality selector and to measure the resulting magnetization, chirality, and critical correlations.

The article is organized as follows.  We formulate the general charge-selector protocol in Section~II and develop the finite-interval TBA for the current-carrying sector and analyze the pure $Q_3$ selector in Section~III.  We present the zero-temperature phase diagram of $H_{J\chi h}$ and explain how the selector reaches the XXX trace center in Section~IV. The zero-entropy property, the emergent Luttinger liquid, and transport are discussed in Section~V.  The experimental implementations are discussed in Section~VI.  In the appendices, we give the technical TBA and low-energy derivations.

\section{Nonequilibrium-from-equilibrium protocol}
\label{sec:selector}

We now formulate the general selector construction used throughout this work.  Let
$H_0$ denote the physical Hamiltonian whose eigenstates we wish to classify, and let
$\{Q_i\}_{i=1}^{m}$ be a set of extensive conserved charges satisfying
\begin{equation}
  [H_0,Q_i]=0,\qquad [Q_i,Q_j]=0 .
\end{equation}
Here $i,j=1,\ldots,m$ label the chosen charges, and all extensive quantities are
understood for a chain of length $L$.  Since the charges commute with $H_0$ ($H_0\equiv JQ_2$ for the XXX model considered in this work),
there exists a common eigenbasis \(\{|n\rangle\}\) such that
\begin{equation}
  H_0|n\rangle = E_n^{(0)} |n\rangle,\qquad
  Q_i|n\rangle = q_{i,n}|n\rangle .
\end{equation}
The numbers \(E_n^{(0)}\) and \(q_{i,n}\) are the energy and conserved-charge
quantum numbers of the eigenstate \(|n\rangle\).  In the thermodynamic limit it is
often more useful to work with the corresponding densities
\begin{equation}
  e_n^{(0)}=\frac{E_n^{(0)}}{L},\qquad
  \mathfrak q_{i,n}=\frac{q_{i,n}}{L}.
\end{equation}

The key idea is to introduce an auxiliary Hamiltonian
\begin{equation}
  H(\boldsymbol{\alpha})
  =
  H_0-\boldsymbol{\alpha}\cdot\mathbf Q
  =
  H_0-\sum_{i=1}^{m}\alpha_i Q_i ,
  \label{eq:selectorHamiltonian}
\end{equation}
where
$\boldsymbol{\alpha}=(\alpha_1,\ldots,\alpha_m)$ is a set of fields conjugate
to the conserved charges and
$\mathbf Q=(Q_1,\ldots,Q_m)$.  Because all $Q_i$ commute with $H_0$, the
Hamiltonian $H(\boldsymbol{\alpha})$ has the same eigenvectors as $H_0$,
with only the ordering of their eigenvalues changed:
\begin{equation}
  H(\boldsymbol{\alpha})|n\rangle
  =
  \left(
    E_n^{(0)}-\sum_i\alpha_i q_{i,n}
  \right)|n\rangle .
\end{equation}
Thus \(H(\boldsymbol{\alpha})\) does not create new eigenstates of the original
problem.  Instead, it acts as a deterministic selector: by favoring particular
charge quantum numbers, it exposes rare simultaneous eigenstates of \(H_0\).

At finite temperature \(T\), with \(\beta=1/T\), the tilted partition function is
\begin{equation}
  Z(\beta,\boldsymbol{\alpha})
  =
  \mathrm{Tr}\,
  e^{-\beta H(\boldsymbol{\alpha})}
  =
  \mathrm{Tr}\,
  e^{-\beta(H_0-\boldsymbol{\alpha}\cdot\mathbf Q)} ,
  \label{eq:tiltedPartitionFunction}
\end{equation}
and the corresponding free energy is
\begin{equation}
  F(\beta,\boldsymbol{\alpha})
  =
  -\beta^{-1}\ln Z(\beta,\boldsymbol{\alpha}) .
\end{equation}
In the common eigenbasis of \(H_0\) and \(Q_i\), the statistical weight of the
state \(|n\rangle\) is
\begin{equation}
  p_n(\beta,\boldsymbol{\alpha})
  =
  \frac{
    \exp\!\left[-\beta\left(E_n^{(0)}
    -\sum_i\alpha_i q_{i,n}\right)\right]
  }{
    Z(\beta,\boldsymbol{\alpha})
  } .
\end{equation}
Equivalently, in terms of densities,
\begin{equation}
  p_n(\beta,\boldsymbol{\alpha})
  \propto
  \exp\!\left[
    -\beta L
    \left(
      e_n^{(0)}
      -\sum_i\alpha_i\mathfrak q_{i,n}
    \right)
  \right] .
\end{equation}
The fields \(\alpha_i\) therefore implement a large-deviation bias in the space
of conserved-charge densities.  At finite temperature this construction is the
usual generalized Gibbs ensemble (GGE) logic: one fixes the average values of
commuting charges by coupling them to conjugate generalized chemical potentials.
The resulting ensemble is generally dominated by a finite-entropy set of Bethe
configurations, or equivalently by a thermodynamic Bethe macrostate with smooth
root and hole densities~\cite{rigol2007,CassidyClarkRigol2011,Yuzbashyan2016,vidmar2016,ilievski2016,BertiniMFT2015}.

The response of the tilted ensemble to the selector fields is obtained by
differentiating the free energy.  One finds
\begin{equation}
  \frac{\partial F}{\partial \alpha_i}
  =
  -\langle Q_i\rangle_{\beta,\boldsymbol{\alpha}},
  \label{eq:FDerivativeCharge}
\end{equation}
where \(\langle\cdots\rangle_{\beta,\boldsymbol{\alpha}}\) denotes the thermal
average with respect to \(H(\boldsymbol{\alpha})\).  A second derivative gives
the charge susceptibility matrix,
\begin{equation}
  \frac{\partial \langle Q_i\rangle_{\beta,\boldsymbol{\alpha}}}
       {\partial \alpha_j}
  =
  \beta\,
  \mathrm{Cov}_{\beta,\boldsymbol{\alpha}}(Q_i,Q_j),
  \label{eq:chargeSusceptibility}
\end{equation}
with covariance
\begin{equation}
  \mathrm{Cov}_{\beta,\boldsymbol{\alpha}}(A,B)
  =
  \langle AB\rangle_{\beta,\boldsymbol{\alpha}}
  -
  \langle A\rangle_{\beta,\boldsymbol{\alpha}}
  \langle B\rangle_{\beta,\boldsymbol{\alpha}} .
\end{equation}
These relations show explicitly that \(\alpha_i\) controls the expectation value
of \(Q_i\), while the covariance matrix measures the fluctuations of conserved
charges in the selected ensemble.

The zero-temperature limit is the central one for the present work.  As
\(\beta\to\infty\), the tilted ensemble collapses onto the ground state of
\(H(\boldsymbol{\alpha})\).  Since \(H(\boldsymbol{\alpha})\) is diagonal in the
same basis as \(H_0\), its ground-state energy is
\begin{equation}
  E_0(\boldsymbol{\alpha})
  =
  \min_n
  \left\{
    E_n^{(0)}
    -
    \sum_i \alpha_i q_{i,n}
  \right\}.
  \label{eq:LFextensive}
\end{equation}
Equivalently, after dividing by \(L\),
\begin{equation}
  e_0(\boldsymbol{\alpha})
  =
  \min_n
  \left\{
    e_n^{(0)}
    -
    \sum_i \alpha_i \mathfrak q_{i,n}
  \right\}.
  \label{eq:LFdensity}
\end{equation}
This is the Legendre--Fenchel transform of the joint spectrum of the conserved
charge densities.  The distinction from an ordinary Legendre transform is
important: if the joint spectrum has flat faces or nonanalytic boundaries, a
single value of \(\boldsymbol{\alpha}\) may select a degenerate set of extremal
states rather than a unique state.  Such nonanalyticities are the spectral
counterpart of first-order transitions, symmetry-sector selection, or spontaneous
symmetry breaking in the thermodynamic limit.

When the minimizing state is unique and the function \(E_0(\boldsymbol{\alpha})\)
is differentiable, the Hellmann--Feynman theorem gives
\begin{equation}
  \frac{\partial E_0}{\partial \alpha_i}
  =
  -\langle Q_i\rangle_{\mathrm{GS}(\boldsymbol{\alpha})},
  \label{eq:HFextensive}
\end{equation}
or, in density form,
\begin{equation}
  \frac{\partial e_0}{\partial \alpha_i}
  =
  -\mathfrak q_i^{\mathrm{GS}}(\boldsymbol{\alpha}) .
  \label{eq:HFdensity}
\end{equation}
Here
\(\ket{\mathrm{GS}(\boldsymbol{\alpha})}\) denotes the ground state of the
selector Hamiltonian \(H(\boldsymbol{\alpha})\), and
\(\mathfrak q_i^{\mathrm{GS}}=\langle Q_i\rangle_{\mathrm{GS}}/L\) is the
corresponding charge density.  Thus the slope of the selector ground-state
energy directly gives the conserved-charge densities of the selected eigenstate
of \(H_0\).

Geometrically, \(\boldsymbol{\alpha}\) defines a supporting hyperplane of the
joint spectrum of
\begin{equation}
  (H_0,Q_1,\ldots,Q_m).
\end{equation}
The zero-temperature selector chooses the exposed point, or exposed face, that
minimizes
\begin{equation}
  E^{(0)}-\boldsymbol{\alpha}\cdot\mathbf q .
\end{equation}
This is why the construction is naturally interpreted as a microcanonical
large-deviation selector.  It does not heat the original Hamiltonian \(H_0\);
rather, it selects an eigenstate whose conserved-charge quantum numbers are
atypical relative to the finite-entropy manifold dominating ordinary thermal or
GGE ensembles.  In Bethe-ansatz language, the finite-temperature GGE usually
selects a macrostate with nonzero Yang--Yang entropy density, whereas the
zero-temperature selector can expose an extremal Bethe macrostate with sharp
occupation functions and vanishing Yang--Yang entropy density.

This point is essential for the physics addressed here.  The auxiliary
Hamiltonian \(H(\boldsymbol{\alpha})\) is easier to analyze because it can be
treated as an equilibrium ground-state problem.  Nevertheless, the resulting
ground state is simultaneously an exact eigenstate of the original Hamiltonian
\(H_0\).  Hence equilibrium methods---Bethe ansatz, conformal field theory,
tensor networks, Monte Carlo, or other ground-state techniques---can be used to
construct and diagnose highly excited, atypical eigenstates of \(H_0\).  The
procedure is therefore ``nonequilibrium from equilibrium'': the nonequilibrium
state of the physical Hamiltonian is obtained as the equilibrium ground state of
a charge-tilted selector.

The same logic remains useful when the charges are not exactly conserved but are
long-lived.  If weak integrability-breaking perturbations, Floquet driving, or
other microscopic effects make the \(Q_i\) only approximately conserved, then
\(H(\boldsymbol{\alpha})\) no longer selects exact eigenstates of \(H_0\).
Instead, it prepares prethermal sectors whose local properties track those of
the charge-selected state over times shorter than the relaxation time associated
with charge decay or Floquet heating~\cite{abanin2015prethermal,abanin2017prethermal,mori2018}.
In that setting the selector construction gives a controlled description of
long-lived nonequilibrium regimes rather than exact stationary eigenstates.

From an experimental viewpoint, the fields \(\alpha_i\) can be interpreted as
static couplings to local or quasilocal charge densities.  Closely related
charge-generated deformations that preserve integrability have been studied in
long-range spin chains and in \(T\bar T\)-type flows
\cite{bargheer2009,smirnov2017ttbar,longrangeTTbar}.  In those works the
deformed Hamiltonian is usually regarded as the physical Hamiltonian itself.
Here the emphasis is different: \(H_0\) remains the physical XXX Hamiltonian,
while \(H(\boldsymbol{\alpha})\) is used only as a selector of special
eigenstates already present in the spectrum of \(H_0\).  Related large-deviation
tilts of conserved charges and atypical generalized ensembles in integrable
chains have also been explored in Refs.~\cite{cecile2024,krajnik2024}.  A
complementary nonequilibrium-from-equilibrium route based on
Knizhnik--Zamolodchikov equations, boundary Wess--Zumino--Novikov--Witten
structures, and other exact integrable time-dependent constructions appears in
Refs.~\cite{SedrakyanBabujian2022,SedrakyanGalitski2010,Yuzbashyan2018AOP}.

In the remainder of this article we apply this general protocol to the
spin-$s=\tfrac12$ isotropic Heisenberg XXX chain.  We use the parity-odd charge
\(Q_3\), together with the spin charge \(S^z\), to select ordered,
current-carrying, critical, zero-entropy Bethe macrostates at finite XXX energy
density.  This provides an exactly solvable realization of atypical eigenstates
whose conserved-charge quantum numbers place them outside the finite-entropy
Bethe manifold that dominates thermal and generalized Gibbs ensembles.


\section{Charge-selected chiral current-carrying sector in the Heisenberg chain}
\label{sec:benchmark-corrected}

We now specialize the general selector protocol of Sec.~\ref{sec:selector} to the spin-$\tfrac12$ isotropic Heisenberg chain.  The selector Hamiltonian is
\begin{equation}
  H_{J\chi h}=JQ_2+\alpha Q_3-hS^z,
\end{equation}
where $Q_2$ is the nearest-neighbor XXX exchange charge defined in Eq.~\eqref{eq:HXXX}, $Q_3$ is the parity- and time-reversal-odd scalar-chirality charge defined in Eq.~\eqref{eq:Q3-intro}, $S^z=\sum_jS_j^z$ is the conserved total magnetization, and $J$, $\alpha$, and $h$ are the corresponding conjugate fields.  The important algebraic point is that $Q_2$, $Q_3$, and $S^z$ commute.  Therefore changing $J$, $\alpha$, or $h$ does not create new eigenvectors; it changes which simultaneous XXX eigenstate is lowest in the tilted spectrum.  In this section we compute that selected eigenstate when the tilt favors a chiral current.

The charge $Q_3$ is the first local charge in the XXX hierarchy that distinguishes the two spatial orientations.  It therefore acts as a handedness selector: for one sign of $\alpha$ it lowers the energy of right-moving spin flips, while for the opposite sign it lowers the energy of left-moving spin flips.  At finite density this one-particle bias becomes an asymmetric Bethe sea.  Here, we proceed in three steps.  First, the one-magnon limit fixes the sign convention and the saturation edge.  Second, the zero-temperature finite-interval TBA gives the dressed energy and root density of the chiral Bethe sea.  Third, the pure current selector $H_{\chi h}=\alpha Q_3-hS^z$ is used as the minimal example of a magnetochiral, current-carrying XXX eigenstate.

\subsection{One-magnon orientation and saturation edge}

Let
$ \ket{F}=\ket{\uparrow\uparrow\cdots\uparrow}
$
be the fully polarized ferromagnetic state on a periodic chain of length $L$.  A one-magnon state is a single spin flip on this background,
\begin{equation}
  \ket{k}=\frac{1}{\sqrt L}\sum_{j=1}^{L}e^{\ii kj}S_j^-\ket{F},
  \qquad k=\frac{2\pi n}{L},\quad n\in\mathbb Z .
\end{equation}
We use the rapidity variable $x$ related to the momentum $k$ by
\begin{equation}
  e^{\ii k}=\frac{x+\ii}{x-\ii},
  \qquad x=\cot\frac{k}{2} .
  \label{eq:rapidity-map}
\end{equation}
Thus
\begin{equation}
  \cos k=\frac{x^2-1}{x^2+1},
  \qquad
  \sin k=\frac{2x}{x^2+1} .
\end{equation}
With the scalar-chirality orientation fixed in Eq.~\eqref{eq:Q3-intro}, the one-magnon charge eigenvalues, measured relative to the fully polarized background for $Q_2$, are
\begin{equation}
  q_2(k)=\cos k-1,
  \qquad
  q_3(k)=-\sin k\,(1-\cos k).
  \label{eq:magnon-dispersion}
\end{equation}
The short lattice derivation of Eq.~\eqref{eq:magnon-dispersion} is given in Appendix~\ref{app:one-magnon-charges}.  In rapidity variables this becomes
\begin{equation}
  q_2^{(1)}(x)=-\frac{2}{1+x^2},
  \qquad
  q_3^{(1)}(x)=-\frac{4x}{(1+x^2)^2} .
  \label{eq:one-string-bare-charges}
\end{equation}
The superscript $(1)$ indicates a Bethe $1$-string.  The sign of $q_3^{(1)}$ fixes the handedness used throughout the paper.  Reversing the definition of the scalar chirality, or using the opposite plane-wave convention $e^{-\ii kj}$, sends $q_3\to -q_3$ and can be absorbed by $\alpha\to-\alpha$. The phase structure remains unchanged.

A one-magnon excitation above $\ket{F}$ lowers $S^z$ by one unit.  Its energy relative to the fully polarized state in the selector Hamiltonian is therefore
\begin{equation}
  \omega_{J\chi h}(k)
  =h+J\bigl(\cos k-1\bigr)-\alpha\sin k\,(1-\cos k).
  \label{eq:one-magnon-selector}
\end{equation}
The even term proportional to $J$ is the ordinary XXX magnon dispersion, while the odd term proportional to $\alpha$ is the chiral current bias.  This already shows the mechanism that persists at finite density: $Q_3$ does not merely shift all momenta uniformly, but energetically favors one direction of propagation.

For the pure current selector, $J=0$ and $\alpha>0$, the function $q_3(k)$ is minimized at
\begin{equation}
  k_0=\frac{2\pi}{3},
  \qquad
  x_0=\cot\frac{k_0}{2}=\frac{1}{\sqrt3},
  \qquad
  q_3(k_0)=-\frac{3\sqrt3}{4} .
\end{equation}
The fully polarized state is stable as long as $\omega_{0\chi h}(k)\ge0$ for all $k$.  Hence the saturation field is
\begin{equation}
  h_c=\frac{3\sqrt3}{4}\,\alpha .
  \label{eq:hc-corrected}
\end{equation}
At $h=h_c$ the interacting Bethe sea shrinks to the one-magnon point $x_0$.  The same one-magnon reasoning also fixes the lower zero-field boundary of the generalized selector: at $h=0$ the ferromagnetic state is stable for $J/\alpha<-1$, while at $J/\alpha=-1$ a chiral magnon becomes soft.  The quantum phase transition at the upper boundary is not a dilute-magnon instability. It occurs when the finite chiral interval merges with the antiferromagnetic XXX sea.  That finite-density boundary is obtained from the TBA at $J/\alpha=\pi/2$ in Sec.~\ref{sec:phase-diagram}.

\subsection{Finite-interval dressed energy and root density}

We now pass from one magnon to the thermodynamic Bethe macrostate selected by $H_{J\chi h}$.  In the magnetochiral branch relevant here the occupied Bethe roots are $1$-strings whose rapidities fill one finite interval
\begin{equation}
  \mathcal I=[x_-,x_+] .
\end{equation}
The endpoints are not imposed by the zeros of the bare driving term.  They are dressed Fermi edges fixed self-consistently by the many-body Bethe sea.

We use the standard XXX kernels within the TBA framework:
\begin{equation}
  a_n(x)=\frac{1}{\pi}\frac{n}{x^2+n^2},
  \qquad n=1,2,\ldots .
  \label{eq:akernel}
\end{equation}
The bare one-string charges in Eq.~\eqref{eq:one-string-bare-charges} may be written as
\begin{equation}
  q_2^{(1)}(x)=-2\pi a_1(x)=-\frac{2}{x^2+1},
  \label{eq:q2bare}
\end{equation}
\begin{equation}
  q_3^{(1)}(x)=2\pi a_1'(x)=-\frac{4x}{(x^2+1)^2}.
  \label{eq:q3bare}
\end{equation}
The bare selector driving term for a $1$-string is
\begin{equation}
  d_{J,h}(x)=h+Jq_2^{(1)}(x)+\alpha q_3^{(1)}(x).
  \label{eq:bare-selector-driving}
\end{equation}
It is the energy of an isolated rapidity before the backflow of all other rapidities has been included.

The zero-temperature finite-interval TBA dresses this driving term into a quasiparticle energy $\varepsilon(x)$ satisfying
\begin{equation}
  \varepsilon(x)+\int_{\mathcal I}dy\,a_2(x-y)\varepsilon(y)=d_{J,h}(x),
  \qquad x\in\mathcal I .
  \label{eq:eps1}
\end{equation}
The filled region is determined by
\begin{equation}
  \varepsilon(x_-)=\varepsilon(x_+)=0,
  \qquad
  \varepsilon(x)<0\quad (x\in\mathcal I).
  \label{eq:Fpoints}
\end{equation}
Thus $\varepsilon(x)$ is a dressed Landau quasiparticle energy measured relative to the selected Bethe sea, and $x_\pm$ are its two Fermi edges.  This is the main finite-density difference between the full TBA and the one-magnon discussion: an added root scatters from all occupied roots and produces Bethe backflow.  Consequently the occupied interval is an interacting Fermi domain, not the set where the bare function $d_{J,h}(x)$ is negative.

The same interval dresses the density of occupied roots.  The root density $\rho(x)$ obeys
\begin{equation}
  \rho(x)+\int_{\mathcal I}dy\,a_2(x-y)\rho(y)=a_1(x),
  \qquad x\in\mathcal I .
  \label{eq:rho-corrected-secIII}
\end{equation}
Equations~\eqref{eq:eps1}--\eqref{eq:rho-corrected-secIII} define one thermodynamic Bethe macrostate.  All observables must be computed from this same solution.  This point is physically important: the chiral field $\alpha$ first biases rapidities asymmetrically, but the kernel then redistributes the whole sea.  Magnetization, chirality, XXX energy density, and low-energy velocities are therefore collective dressed properties of one sharp macrostate.

The selector free-energy density, including the fully polarized background contribution, can be written as
\begin{equation}
  f_{J\chi h}=\frac{J}{4}-\frac{h}{2}
  +\int_{\mathcal I}dx\,\rho(x)d_{J,h}(x).
  \label{eq:f-density}
\end{equation}
Equivalently,
\begin{equation}
  f_{J\chi h}=\frac{J}{4}-\frac{h}{2}
  +\int_{\mathcal I}dx\,a_1(x)\varepsilon(x).
  \label{eq:f}
\end{equation}
The equality follows from the self-adjointness of the finite-interval integral operator and is derived in Appendix~\ref{app:tba}.  Differentiating $f_{J\chi h}$ gives the order parameters.  Since $\varepsilon(x_\pm)=0$, endpoint terms vanish, and the Hellmann--Feynman relations give
\begin{equation}
  M=-\partial_h f_{J\chi h}
  =\frac12-\int_{\mathcal I}dx\,\rho(x),
  \label{eq:M}
\end{equation}
where $M=\langle S^z\rangle/L$ is the magnetization density, and
\begin{equation}
  \chi=\partial_\alpha f_{J\chi h}
  =\int_{\mathcal I}dx\,\rho(x)q_3^{(1)}(x),
  \label{eq:chi}
\end{equation}
where $\chi=\langle Q_3\rangle/L$ is the scalar-chirality density in the convention of Eq.~\eqref{eq:Q3-intro}.  The energy density of the same state under the undeformed XXX Hamiltonian is
\begin{equation}
  e_{\rm XXX}=\frac14+
  \int_{\mathcal I}dx\,\rho(x)q_2^{(1)}(x).
  \label{eq:eXXX-general}
\end{equation}
The one-string description is stable against finite-rapidity higher-string condensation throughout the magnetochiral branch; the explicit test pseudoenergies are given in Appendix~\ref{app:higher-string-stability}.  The only zero mode that remains at $h=0$ is the expected global $SU(2)$ spin-rotation degeneracy.

This finite-interval formulation also makes the large-deviation character of the selected state explicit.  A finite-temperature GGE is dominated by smooth particle and hole occupations and has nonzero Yang--Yang entropy.  Here the zero-temperature selector exposes an extremal Bethe macrostate with a sharp occupation domain.  The two Fermi edges are nevertheless gapless, so the state is both zero-entropy and critical.  The Luttinger liquid discussed in Sec.~\ref{sec:critical} is precisely the theory of particle-hole fluctuations around these two dressed edges.

\subsection{Pure current selector and zero-field magnetochiral state}

We first set $J=0$, so that
\begin{equation}
  H_{\chi h}=\alpha Q_3-hS^z .
\end{equation}
For $|h|\ge h_c$ the selector ground state is fully polarized.  For $0<|h|<h_c$ the TBA interval is open and describes a chiral critical Bethe sea.  The branch with $h<0$ is obtained by spin reversal: $M(\alpha,-h)=-M(\alpha,h)$, while $\chi(\alpha,-h)=\chi(\alpha,h)$.

At $h=0^+$, solving the TBA integral Eqs.~\eqref{eq:eps1}--\eqref{eq:Fpoints} gives
\begin{equation}
  x_-=0.04301748\ldots,
  \qquad
  x_+=3.40164754\ldots .
  \label{eq:hzero-endpoints-corrected}
\end{equation}
The density equation on this interval gives
\begin{align}
  M(0^\pm)&=\pm0.21563609\ldots,
  \nonumber\\
  \chi(0)&=-0.24515023\ldots .
  \label{eq:zero}
\end{align}
Thus the pure current selector produces a magnetochiral state already at zero external field, in the one-sided thermodynamic sense.  The chirality is finite because the Bethe sea is strongly asymmetric in rapidity space.  The magnetization is finite because of the spontaneous symmetry breaking: the selected Bethe state belongs to an $SU(2)$ multiplet whose total spin is extensive.

Let $N$ be the number of Bethe roots.  A regular finite-chain XXX Bethe state is a highest-weight state of the global $SU(2)$ algebra: $S^+\ket{\{x_j\}}=0$, $S^z=L/2-N$, and $S=L/2-N$~\cite{takahashi_book,korepin_book}.  In the thermodynamic finite-interval state,
\begin{equation}
  \frac{N}{L}=\int_{\mathcal I}dx\,\rho(x),
  \qquad
  \frac{S}{L}=\frac12-\int_{\mathcal I}dx\,\rho(x)=M(0^+).
  \label{eq:spin-density-highest-weight}
\end{equation}
The $h\to0^+$ limit therefore selects the highest-weight representative of a giant multiplet, and the $h\to0^-$ limit selects the lowest-weight representative.  This is the finite-volume mechanism by which the thermodynamic order parameter appears.  The nontrivial statement is that the charge-selected multiplet has $S/L=0.215636\ldots$, whereas in the antiferromagnetic-descended no-SSB branch the same Bethe counting gives $S/L\to0$.  A small longitudinal field chooses a direction inside an already extensive multiplet, exactly as a symmetry-breaking field chooses a representative of a thermodynamic ordered phase.

\begin{figure*}[t]
    \centering
    \includegraphics[width=0.86\textwidth]{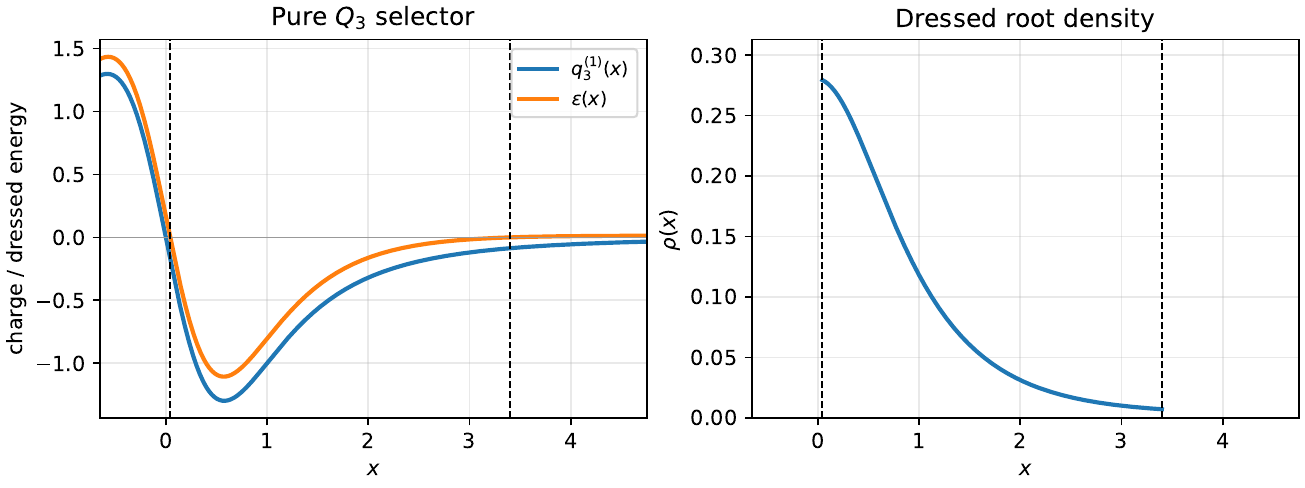}
    \caption{Zero-field finite-interval TBA solution for the pure local $Q_3$ selector, $J=0$, $h=0^+$, and $\alpha=1$.  Left: bare scalar-chirality charge $q_3^{(1)}(x)$ and dressed energy $\varepsilon(x)$.  The dashed vertical lines are the dressed Fermi edges $x_-$ and $x_+$.  The dressed interval is determined by $\varepsilon(x_\pm)=0$, not by the bare charge alone.  Right: dressed root density $\rho(x)$ on the same interval.  The asymmetric distribution of occupied roots gives finite chirality and finite one-sided magnetization.}
    \label{fig:pureQ3profile}
\end{figure*}

In Figure~\ref{fig:pureQ3profile} we show the finite-interval solution for $J=0$, $h=0^+$, and $\alpha=1$.  The left panel compares the bare chirality charge $q_3^{(1)}(x)$ with the dressed energy $\varepsilon(x)$.  The dressed Fermi edges are fixed by $\varepsilon(x_\pm)=0$, not just by the bare charge.  The right panel shows the corresponding root density.  Its asymmetric support and weight distribution are the microscopic origin of the finite scalar chirality.

The zero-field pure-$Q_3$ state is also an eigenstate of the undeformed XXX Hamiltonian.  Its XXX energy density is
\begin{equation}
  e_{\rm XXX}^{\chi}
  =-0.12396686\ldots .
  \label{eq:xxx-energy-hzero}
\end{equation}
This value lies above the antiferromagnetic ground-state density $1/4-\ln2$ and below the full Hilbert-space trace center $e_{\rm XXX}=0$.  Hence the minimal current selector already produces a high-energy, current-carrying, zero-entropy eigenstate, although at $h=0^+$ it is not exactly at the trace center.

The trace center can be reached within the pure current selector by keeping $J=0$ and increasing the one-sided field.  Define $h_*$ by
\begin{equation}
  e_{\rm XXX}(h_*)=0,
  \label{eq:hstar-definition}
\end{equation}
where $e_{\rm XXX}(h)$ is computed from Eq.~\eqref{eq:eXXX-general} on the pure-$Q_3$ branch.  The finite-interval TBA gives
\begin{align}
  h_*/h_c&=0.62865083\ldots,
  \nonumber\\
  M_*&=0.32686911\ldots,
  \nonumber\\
  \chi_*&=-0.19822567\ldots .
  \label{eq:hstar-params}
\end{align}
This trace-center state is magnetized and chiral, but the field $h_*$ explicitly selects the spin sector.  In Sec.~\ref{sec:phase-diagram} we show that tuning the commuting exchange bias $JQ_2$ provides a complementary route to the trace center in the one-sided $h\to0^\pm$ limit.

\begin{figure*}[t]
    \centering
    \includegraphics[width=0.86\textwidth]{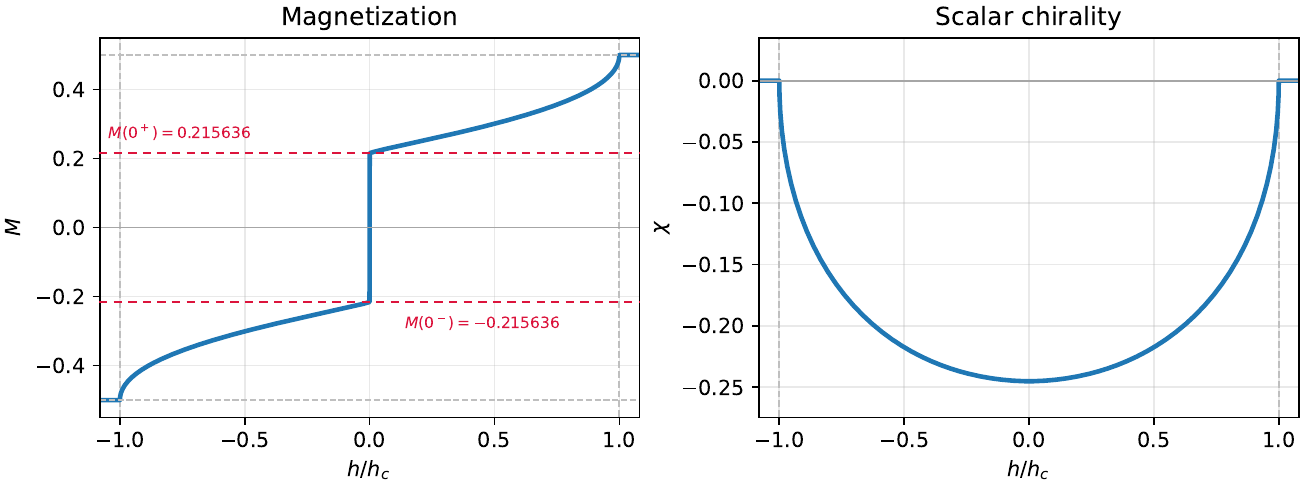}
    \caption{Magnetization density $M$ and scalar-chirality density $\chi$ for the pure current selector $H_{\chi h}=\alpha Q_3-hS^z$, obtained from the finite-interval TBA.  The horizontal axis is normalized by the saturation field $h_c=3\sqrt3\alpha/4$.  The range extends slightly beyond $|h|/h_c=1$ to show the fully polarized region, where $|M|=1/2$ and $\chi=0$.  The red dashed horizontal lines in the left panel mark the spontaneous one-sided magnetization values $M(0^+)=0.215636\ldots$ and $M(0^-)=-0.215636\ldots$.  These numbers equal $\pm S/L$ for the highest- and lowest-weight representatives of the macroscopic zero-field $SU(2)$ multiplet.  The jump of $M$ at $h=0$ is the thermodynamic one-sided limit in which an infinitesimal field selects the orientation of this giant multiplet.}
    \label{fig:MChi}
\end{figure*}

In Figure~\ref{fig:MChi} we show the full field dependence of the pure current selector.  The magnetization exhibits the one-sided jump in Eq.~\eqref{eq:zero}, while the chirality evolves smoothly and vanishes at saturation.  Close to $|h|=h_c$, the occupied interval becomes a dilute pocket around $x_0=1/\sqrt3$, and one obtains
\begin{align}
  \frac12-|M| &= \frac{1}{\pi}
  \sqrt{1-\frac{|h|}{h_c}}
  +O\!\left(1-\frac{|h|}{h_c}\right), \nonumber\\
  |\chi| &= \frac{3\sqrt3}{4\pi}
  \sqrt{1-\frac{|h|}{h_c}}
  +O\!\left(1-\frac{|h|}{h_c}\right).
  \label{eq:saturationcheck}
\end{align}
The square-root behavior is the standard closing of a one-dimensional Fermi pocket.  It gives a useful analytic check on the finite-interval numerics and connects the interacting TBA solution continuously to the one-magnon saturation edge.

The pure-selector analysis identifies the local current-carrying eigenstate, its saturation edge, and its one-sided order parameters.  Below, we restore the commuting exchange bias $JQ_2$.  This does not change the eigenbasis, but it moves the selected Bethe macrostate through the undeformed XXX energy band.  The next section shows that the magnetochiral branch then spans the entire thermodynamic XXX energy range and includes a zero-field route to the trace center.

\section{Zero-temperature phase diagram of \texorpdfstring{$H_{J\chi h}$}{H J chi h}}
\label{sec:phase-diagram}

The phase diagram of Eq.~\eqref{eq:HJchih-intro} organizes the selected eigenstates by two knobs.  The chiral field $\alpha$ chooses the direction and strength of the current bias, while $J$ shifts the selected state along the XXX energy axis.  Since $JQ_2$ commutes with $Q_3$ and $S^z$, this tuning does not mix eigenstates or break integrability.  It changes the supporting hyperplane in charge space, and hence changes which simultaneous eigenstate of the XXX hierarchy is exposed.  We first show how the zero-field branch reaches the full trace center, then describe the order-parameter window, and finally summarize the full $(J/\alpha,h/\alpha)$ phase diagram.

At $h=0^+$ and $\alpha=1$, solving Eq.~\eqref{eq:eXXX-general} with $e_{\rm XXX}=0$ gives
\begin{align}
  \frac{J_0}{\alpha}&=-0.52207095\ldots,\nonumber\\
  M_0&=0.28121525\ldots,\nonumber\\
  \chi_0&=-0.21188188\ldots .
  \label{eq:center-params}
\end{align}
These values confirm that by adding a negative $JQ_2$ bias one selects a sharp magnetochiral state exactly at the trace center of the pure XXX spectrum, while retaining finite magnetization and finite scalar chirality.  The unrestricted thermal ensemble of the undeformed XXX chain at this full trace-center energy density would be the infinite-temperature ensemble, with no chirality, no preferred spin direction, and maximal entropy.  The charge-selected state is different because its Bethe occupations remain sharp: it is an ordered, critical, zero-entropy Bethe macrostate at the same XXX energy density as an otherwise featureless thermal state.

The finite-interval magnetochiral branch exists for
\begin{equation}
  -1 < J/\alpha < \pi/2 .
  \label{eq:J-magnetochiral-window}
\end{equation}
The interval is not symmetric in $J/\alpha$.  The lower boundary, $J/\alpha=-1$, is the one-magnon instability of the fully polarized ferromagnet discussed above.  The upper boundary, $J/\alpha=\pi/2$, is the chiralization threshold of the antiferromagnetic XXX sea, equivalently $\alpha_c/J=2/\pi$ on the positive-$J$ side~\cite{WMS2024}.  Inside the interval in Eq.~\eqref{eq:J-magnetochiral-window}, an infinitesimal field selects a thermodynamic state with finite $M$ and finite $\chi$; this is the magnetochiral symmetry-broken regime.  For $J/\alpha>\pi/2$ the selector ground state is continuously connected to the antiferromagnetic XXX sea; it is a no-SSB XY-paramagnetic critical sector with no scalar chirality and no spontaneous spin polarization at $h\to0^\pm$.  For $J/\alpha<-1$ it is the fully polarized FM XXX selector state with $\chi=0$; it is therefore outside the magnetochiral high-energy sector studied here.

The zero-field order-parameter scan introduced in Figure~\ref{fig:Jorders} makes this window explicit.  The same green branch, $-1<J/\alpha<\pi/2$, is both chiral and one-sided magnetized; hence the spontaneous spin-sector selection occurs throughout the chiral selector ground-state regime rather than only at the pure $Q_3$ point.  The dotted trace-center point lies well inside this interval, where both order parameters are finite.

In Figure~\ref{fig:energy-alphaoverJ}, introduced in the Introduction, we plot the XXX energy density $e_{\rm XXX}$ of the selected state as a function of the inverse tuning ratio $\alpha/J$, at $h=0^+$.  The gray horizontal band shows the thermodynamic energy band of the pure XXX chain, bounded by the Bethe-ansatz antiferromagnetic ground-state density
\begin{equation}
  e_{\rm AF}=1/4-\ln2
\end{equation}
and by the fully polarized maximum $e_{\rm F}=1/4$.  The dotted horizontal line is the full Hilbert-space trace center $e_{\rm XXX}=0$, which is the unrestricted infinite-temperature energy density of the pure XXX chain.  This is distinct from the center of a symmetry-resolved sector at fixed nonzero magnetization.  The curve is obtained by solving Eqs.~\eqref{eq:eps1}--\eqref{eq:eXXX-general} for each value of $J/\alpha$ and then replotting the answer versus $\alpha/J$.  On both sides of $J=0$, the magnetochiral curve approaches the same pure-current value $e_{Q_3}=e_{\rm XXX}^{\chi}=-0.12396686\ldots$, shown by the blue dash-dotted line.  The positive-$J$ critical point is marked at
\begin{equation}
  \alpha_c/J=2/\pi .
  \label{eq:alpha-critical-over-J}
\end{equation}
The zero-temperature transition should be separated from the finite-temperature response of the nonchiral side.  In the antiferromagnetic-descended region $0<\alpha/J<2/\pi$, the $T=0$ Bethe sea is symmetric and its scalar chirality is exactly zero.  Finite temperature, however, populates particle-hole excitations around both Fermi points.  Because the $Q_3$ charge is odd under spatial reversal, the chiral bias changes the two branch velocities in opposite directions.

The finite-temperature response of the nonchiral side is distinct from the zero-temperature magnetochiral branch.  At low temperature, the two chiral components of the $c=1$ Luttinger liquid contribute
\begin{equation}
  f_T(\alpha)-f_0(\alpha)
  =
  -\frac{\pi T^2}{12}
  \left[
    \frac{1}{|v_R(\alpha)|}
    +
    \frac{1}{|v_L(\alpha)|}
  \right]
  +\cdots .
  \label{eq:thermal-free-energy-derivation}
\end{equation}
For small chiral bias, parity at $\alpha=0$ gives
$v_R(\alpha)=v_0+u_1\alpha/J+\cdots$ and
$|v_L(\alpha)|=v_0-u_1\alpha/J+\cdots$.  The terms linear in
$\alpha$ therefore cancel in the free energy, and the first nonzero
correction is quadratic in $\alpha$ and proportional to $T^2$.
Differentiating with respect to the chiral bias gives
\begin{equation}
  \chi(T)=\partial_\alpha f_T
  =
  -\kappa\,\frac{\alpha}{J}
  \left(\frac{T}{J}\right)^2
  +\cdots ,
  \qquad
  \kappa>0 ,
  \label{eq:thermal-chirality}
\end{equation}
with the sign fixed by the convention for $Q_3$ in Eq.~\eqref{eq:magnon-dispersion}.  At the isotropic point, subleading corrections may also contain the usual marginal logarithmic renormalizations of the XXX Luttinger liquid.
 This is the finite-temperature chiral response obtained numerically from TBA in the chiral Heisenberg chain~\cite{WMS2024}: it is linear in the chiral bias and quadratic in temperature, but it disappears as $T\to0$.  It is therefore distinct from the magnetochiral branch studied here, where the selected zero-temperature eigenstate has a finite chirality density and, in the one-sided $h\to0^\pm$ limit, a spontaneously chosen spin sector.

Near this point, one should distinguish two energies.  The selector ground-state energy $f_{J\chi0}=J F(\alpha/J)$ has the quadratic onset characteristic of the chiralization transition~\cite{WMS2024}.  By contrast, in Figure~\ref{fig:energy-alphaoverJ} we show the projection of the same state onto $Q_2$, namely $e_{\rm XXX}=\partial f_{J\chi0}/\partial J|_\alpha$.  If $F(\lambda)-F(\lambda_c)$ is quadratic in $\lambda-\lambda_c$, with $\lambda=\alpha/J$, then $e_{\rm XXX}=F-\lambda F'$ has a leading linear term.  Thus the apparent steep onset of the blue curve is not a square-root singularity; it is the expected behavior of the XXX-energy projection.

The magnetochiral branch lies inside the many-body band and reaches the full trace center at
\begin{equation}
  \alpha/J_0=-1.91544848\ldots .
  \label{eq:alpha-over-J0}
\end{equation}
In Figure~\ref{fig:energy-alphaoverJ} we mark this point with the black dot.  Combining the two blue branches in Figure~\ref{fig:energy-alphaoverJ}, one finds magnetochiral symmetry-broken eigenstates throughout the thermodynamic energy band of the undeformed XXX chain: the positive branch covers the interval from the antiferromagnetic edge to the pure-$Q_3$ energy, while the negative branch covers the interval from the pure-$Q_3$ energy to the ferromagnetic edge.  Apart from the limiting endpoints where the chirality vanishes or the state saturates, there is no macroscopic energy sector of the XXX band in which charge-selected magnetochiral eigenstates are absent.  Thus we found that {\it the current-carrying, spin-ordered, zero-entropy eigenstates can be selected not only near one special energy, but across the full thermodynamic band}.

\begin{figure*}[t]
    \centering
    \includegraphics[width=0.88\textwidth]{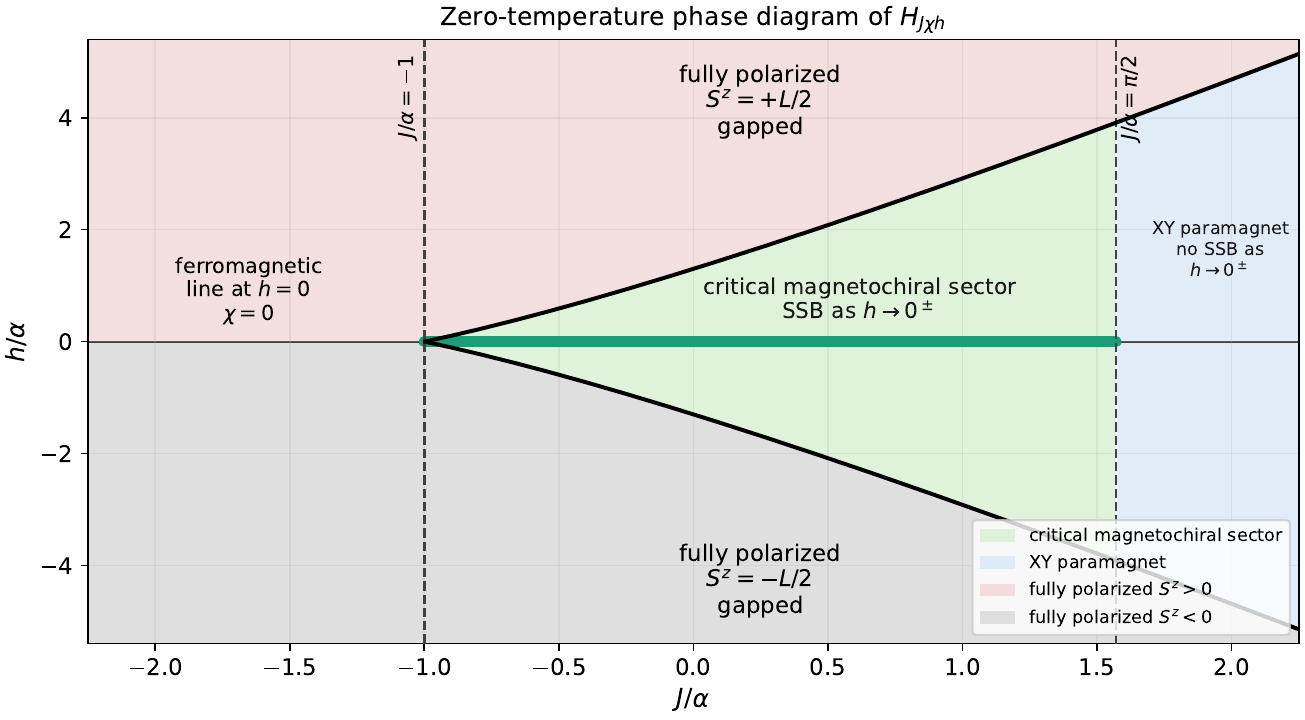}
    \caption{Global zero-temperature phase diagram of the selector $H_{J\chi h}=JQ_2+\alpha Q_3-hS^z$ for $\alpha>0$.  The horizontal axis is the exchange bias $J/\alpha$ and the vertical axis is the field $h/\alpha$.  The black curves are the saturation boundaries where the dressed Fermi interval closes and the selector ground state becomes fully polarized.  Between the black curves the ground state is a one-string critical Bethe sea.  The green region is the critical magnetochiral sector; at $h\to0^\pm$ and $-1<J/\alpha<\pi/2$ it has spontaneous spin-sector selection and finite scalar chirality.  The blue region is an XY-paramagnetic critical sector: a finite field polarizes it smoothly, but it has no spontaneous spin selection as $h\to0^\pm$.  The red and gray regions are fully polarized, gapped states with $S^z=+L/2$ and $S^z=-L/2$, respectively.  The vertical dashed lines mark the ferromagnetic edge $J/\alpha=-1$ and the antiferromagnetic chiralization threshold $J/\alpha=\pi/2$; the highlighted segment on the $h=0$ axis marks the zero-field magnetochiral SSB interval.}
    \label{fig:global-phase}
\end{figure*}
In Figure~\ref{fig:global-phase} we summarize the same solution in the two-dimensional parameter space of the full selector Hamiltonian.  The black curves have a simple one-magnon meaning.  The upper saturation boundary is
\begin{equation}
  h_{\rm sat}^{(+)}(J,\alpha)
  =-\min_k\Big[J(\cos k-1)-\alpha\sin k(1-\cos k)\Big],
  \label{eq:saturation-boundary-main}
\end{equation}
whenever the right-hand side is positive; otherwise the boundary is at zero field.  Spin reversal gives the lower boundary,
\begin{equation}
  h_{\rm sat}^{(-)}(J,\alpha)=-h_{\rm sat}^{(+)}(J,\alpha).
\end{equation}
Above the upper curve a single flipped spin costs positive energy for every momentum, so the all-up ferromagnet is stable and gapped.  Below the lower curve the spin-reversed all-down ferromagnet is stable and gapped.  Between the curves the selector fills a Bethe interval and supports gapless particle-hole excitations.

In Figure~\ref{fig:global-phase} we use the highlighted segment of the $h=0$ axis to distinguish the zero-field magnetochiral interval.  More generally, the zero-field line has three physically different limits.  For $-1<J/\alpha<\pi/2$, the $h\to0^\pm$ limit selects one of two magnetized, chiral sectors.  This is the critical magnetochiral branch: it is gapless, current carrying, and symmetry broken after the thermodynamic one-sided limit is taken.  For $J/\alpha>\pi/2$, the state is continuously connected to the antiferromagnetic XXX sea.  We call this region XY-paramagnetic because it remains critical, but it does not choose a spin direction as $h\to0^\pm$ and has no scalar chirality at zero field.  For $J/\alpha<-1$, the selector is already ferromagnetic at zero field; this is conventional spin order with $\chi=0$, not the high-energy magnetochiral sector.  In this way the phase diagram separates ordinary gapped polarized states, the no-SSB XY-paramagnetic critical sector, and the current-carrying symmetry-broken eigenstates that are the focus of this article.

At $h=0$, both $H_{\chi0}=\alpha Q_3$ and the generalized selector $H_{J\chi0}$ are fully $SU(2)$ invariant: they commute with ${\bf S}_{\rm tot}^2$ and with every component of the total spin.  A finite chain therefore has an exact spin multiplet rather than a unique state with a chosen axis.  The infinitesimal field does not create the macroscopic spin density; it selects the highest- or lowest-weight representative of a multiplet whose total spin has already become extensive in the chiral finite-interval branch.  The phase distinction is therefore the thermodynamic scaling of the multiplet, $S/L>0$ in the magnetochiral branch but $S/L=0$ in the no-SSB XY-paramagnetic branch.  The one-sided limits in Eqs.~\eqref{eq:zero} and~\eqref{eq:center-params} are symmetry-broken limits in exactly this sense: one first takes $L\to\infty$ and only then sends $h\to0^\pm$.  The resulting state selects a magnetization axis and reduces $SU(2)$ to the residual $U(1)$ generated by $S^z$, while at the same time carrying a finite scalar chirality.  The nonrelativistic symmetry-breaking structure is consistent with the standard counting of Nambu--Goldstone modes in systems without Lorentz invariance~\cite{NielsenChadha1976,Leutwyler1994,WatanabeMurayama2012,Hidaka2013}.

The phase diagram thus tells us where the magnetochiral eigenstate exists.  The next question is what kind of many-body state it is: whether it carries entropy, what critical theory describes its soft modes, and how its current-carrying structure appears in transport.

\section{Critical magnetochiral eigenstate: zero entropy, Luttinger liquid, and transport}
\label{sec:critical}
\label{sec:zero-entropy-ll-transport}

The TBA solution above fixes the charge-selected macrostate.  We now explain its physical character.  The state is a finite-energy eigenstate of the undeformed XXX chain, but it is not thermal.  At $h=0^\pm$ and $J=0$ its energy density is given by Eq.~\eqref{eq:xxx-energy-hzero}.  By tuning the commuting energy bias to $J=J_0$ in Eq.~\eqref{eq:center-params}, the selected state reaches the trace-center energy density $e_{\rm XXX}=0$ while keeping finite magnetization and chirality.  In both cases the state is sharply selected by conserved charges rather than sampled from a thermal ensemble.  The finite chirality, one-sided magnetization, and critical correlations therefore coexist with zero thermodynamic entropy density.

\subsection{Sharp Bethe occupations and zero Yang--Yang entropy}
\label{subsec:yy-entropy-main}

For $0<|h|<h_c$, the zero-temperature TBA solution fills one interval of $1$-string rapidities, $\mathcal I=[x_-,x_+]$.  Let $\rho_n(x)$ and $\rho_n^h(x)$ be the particle and hole densities of $n$-strings, and define
\begin{equation}
  \rho_n^t(x)=\rho_n(x)+\rho_n^h(x),
  \qquad
  n_n(x)=\frac{\rho_n(x)}{\rho_n^t(x)} .
  \label{eq:tba_occ_main}
\end{equation}
The simplest exact quantity in the selected macrostate is its Yang--Yang entropy density: it is exactly zero, independent of the numerical solution of the Fermi-edge equations.  The reason is purely kinematic.  The Yang--Yang entropy density of a Bethe macrostate is~\cite{YangYang1969}
\begin{align}
  s_{\rm YY}
  &=\sum_{n\ge1}\int dx\,\rho_n^t(x)\,{\cal S}[n_n(x)],
  \label{eq:YY_main}\\
  {\cal S}(n)&=-n\log n-(1-n)\log(1-n).
  \label{eq:YY_integrand_main}
\end{align}
A partially filled mode with $0<n<1$ carries entropy because many microscopic root configurations realize the same smooth occupation.  A completely full or completely empty mode carries no local mixing entropy.

The selector ground state has precisely such sharp occupations:
\begin{align}
  n_1(x)&=\Theta[-\varepsilon(x)]
  =
  \begin{cases}
    1, & x\in\mathcal I,\\
    0, & x\notin\mathcal I,
  \end{cases}
  \label{eq:n1sharp_main}\\
  n_m(x)&=0,\qquad m\ge2 .\nonumber
\end{align}
The Fermi points have measure zero and do not contribute to the entropy density.  Therefore the integrand in Eq.~\eqref{eq:YY_main} vanishes point by point and
\begin{equation}
  s_{\rm YY}=0 .
  \label{eq:sYYzero_main}
\end{equation}
This is the reason why the magnetochiral state is not thermal.  It may lie deep in the XXX spectrum, and even at the trace center for the energy-tuned selector in Eq.~\eqref{eq:center-params}, but its Bethe distribution is as sharp as a zero-temperature Fermi sea.  It is thus an atypical ordered, critical, zero-entropy Bethe macrostate selected by a large-deviation field.  A finite-temperature GGE with the same average energy density would average over a finite-entropy manifold of Bethe configurations. The charge selector instead exposes a single zero-entropy extremal macrostate outside that manifold.  This is why the state is ETH-violating without invoking disorder and is scar-like but without relying on a finite-dimensional scar subspace.

\subsection{Luttinger liquid from the Fermi edges}
\label{subsec:ll-main}

Zero entropy density does not mean that the state is inert.  The two edges of the filled interval are soft: moving a Bethe root from just inside an edge to just outside it costs arbitrarily little energy in the thermodynamic limit.  These particle-hole excitations form a single compact bosonic mode, i.e., a $c=1$ Luttinger liquid, for each fixed $0\leq|h|<h_c$.  Because the background carries chirality and breaks parity and time reversal, the two branches generally have unequal velocities.

The physical velocities are dressed velocities.  With the density $\rho(x)$ determined by Eq.~\eqref{eq:rho-corrected-secIII}, the dressed momentum satisfies
\begin{equation}
  p'(x)=2\pi\rho(x).
\end{equation}
The Fermi velocities are therefore
\begin{equation}
  v_a=\left.
  \frac{\partial_x\varepsilon(x)}{2\pi\rho(x)}
  \right|_{x=x_a},
  \qquad x_a=x_\pm .
  \label{eq:dressedvelocitymain}
\end{equation}
Here we have exactly the same finite-interval dressed energy and density as the thermodynamics in Sec.~\ref{sec:benchmark-corrected}. 

The Luttinger parameter is obtained from the dressed charge.  After shifting the interval to $[-B,B]$, with $B=(x_+-x_-)/2$, the dressed charge solves
\begin{equation}
  {\cal Z}(t)+\int_{-B}^{B}\!dt'\,a_2(t-t'){\cal Z}(t')=1,
  \label{eq:dressedcharge-main}
\end{equation}
with
\begin{equation}
  K={\cal Z}^2(B).
  \label{eq:K-main}
\end{equation}
The parameter $K$ controls equal-time power-law exponents and the compactification radius, whereas $v_\pm$ control real-time propagation.  This is the chiral, current-carrying analogue of the one-dimensional quantum-liquid phenomenology reviewed in Ref.~\cite{ImambekovSchmidtGlazman2012}.  Thus the static and dynamical signatures of the magnetochiral state are complementary.

\begin{figure*}[t]
    \centering
    \includegraphics[width=0.85\textwidth]{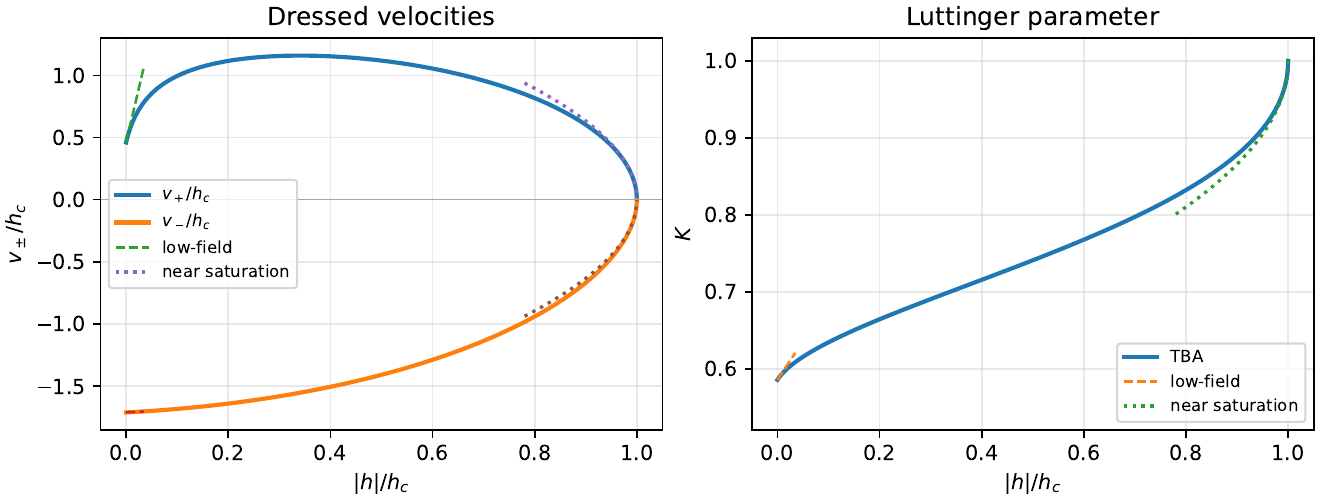}
    \caption{Low-energy data computed from the finite-interval TBA for the pure current selector.  Left: dressed Fermi velocities from Eq.~\eqref{eq:dressedvelocitymain}.  The solid curves are the full TBA results, the dashed curves show the low-field expansion in Eq.~\eqref{eq:lowfield-vK-main}, and the dotted curves show the saturation form in Eq.~\eqref{eq:saturation-vK-main}.  Right: Luttinger parameter from the dressed-charge equation, with the same low-field and saturation asymptotes.  The asymmetric velocities reflect the current-carrying Bethe sea; the approach $K\to1$ near saturation is the dilute-magnon limit.}
    \label{fig:LLdata}
\end{figure*}

The same finite-interval equations also give controlled limiting forms.  For the one-sided $h\to0^+$ branch of the pure current selector,
\begin{align}
  \frac{v_+}{h_c}&=0.46102667+17.46427\frac{|h|}{h_c}
  +O\!\left[(|h|/h_c)^2\right],\nonumber\\
  \frac{v_-}{h_c}&=-1.71169852+0.214981\frac{|h|}{h_c}
  +O\!\left[(|h|/h_c)^2\right],\nonumber\\
  K&=0.58555533+0.998992\frac{|h|}{h_c}
  +O\!\left[(|h|/h_c)^2\right].
  \label{eq:lowfield-vK-main}
\end{align}
For $h\to0^-$ the spin-reversed branch exchanges the two velocities.  Near saturation, the occupied interval becomes a small Fermi pocket and the leading behavior is universal:
\begin{align}
  \frac{v_\pm}{h_c}&=\pm2\sqrt{1-|h|/h_c}
  +O(1-|h|/h_c),\nonumber\\
  K&=1-\frac{4}{3\pi}\sqrt{1-|h|/h_c}
  +O(1-|h|/h_c).
  \label{eq:saturation-vK-main}
\end{align}
The low-field constants are nontrivial dressed quantities, whereas the square-root saturation laws follow from a dilute one-magnon pocket.  In Figure~\ref{fig:LLdata} we show the full finite-interval TBA results for the dressed velocities and Luttinger parameter together with these asymptotic forms.

For fixed nonzero field, the universal finite-size and low-generalized-temperature corrections take the asymmetric-velocity form
\begin{align}
  E_0(L)&=L f(\alpha,h)
  -\frac{\pi}{12L}\big(|v_+|+|v_-|\big)+o(L^{-1}),
  \label{eq:finite-size-cft}\\
  f(T)&=f(0)-\frac{\pi T^2}{12}
  \big(|v_+|^{-1}+|v_-|^{-1}\big)+o(T^2).
  \label{eq:lowT-main}
\end{align}
Here $T$ is the generalized temperature of the selector Hamiltonian, not the physical temperature of the undeformed XXX Hamiltonian.  The equal-time bipartite entanglement of a periodic chain has the standard $c=1$ conformal-field-theory form~\cite{CalabreseCardy2004}
$
  S_\ell=\frac{1}{3}\log\!\left[
  \frac{L}{\pi}\sin\!\left(\frac{\pi\ell}{L}\right)
  \right]+s_0+o(1),
$
with nonuniversal constant $s_0$.  The magnetochiral eigenstate therefore combines zero Yang--Yang entropy with logarithmic critical entanglement.

\subsection{Transverse spin correlations and symmetry breaking}
\label{subsec:transverse-correlations}

The extensive spin of the selected multiplet does not imply a transverse XY ferromagnetic long-range order in the highest-weight representative selected by $h\to0^+$.  Let
\begin{equation}
  C^{+-}(r)=\frac{1}{L}\sum_{j=1}^L
  \Big(\langle S_j^+S_{j+r}^-\rangle
  -\langle S_j^+\rangle\langle S_{j+r}^-\rangle\Big)
  \label{eq:Cpm-def}
\end{equation}
with periodic boundary conditions.  In the $z$-oriented one-sided state, the residual $U(1)$ symmetry gives $\langle S_j^+\rangle=0$, so the connected and ordinary transverse correlators coincide.  For a highest-weight state with total spin $S=ML$, the global $SU(2)$ algebra gives the exact sum rule
\begin{equation}
  \sum_{i,j}\langle S_i^+S_j^-\rangle
  =\langle S_{\rm tot}^+S_{\rm tot}^-\rangle=2S .
  \label{eq:SpSm-sumrule}
\end{equation}
The diagonal part is fixed locally, $\sum_i\langle S_i^+S_i^-\rangle=L(1/2+M)$.  Hence the off-site spatial average obeys
\begin{equation}
  \frac{1}{L(L-1)}\sum_{i\ne j}\langle S_i^+S_j^-\rangle
  =\frac{M-1/2}{L-1}\xrightarrow[L\to\infty]{}0 .
  \label{eq:offsite-average}
\end{equation}
Thus the highest-weight representative has no constant transverse connected correlator.  The finite magnetization is a one-point order parameter selected by the one-sided field, not a transverse two-point plateau.

The nonconstant part of $C^{+-}(r)$ is nevertheless critical.  Bosonizing the two Fermi edges of the finite Bethe interval gives the equal-time asymptotic form
\begin{equation}
  C^{+-}(r)\sim
  \sum_{n\in\mathbb Z} A_n
  \frac{e^{\ii q_n r}}{|r|^{1/(2K)+2n^2K}},
  \qquad r\gg1,
  \label{eq:Cpm-LL}
\end{equation}
where the nonuniversal amplitudes $A_n$ and wave vectors $q_n$ are fixed by the microscopic Bethe sea, while the exponents are fixed by the dressed charge through $K={\cal Z}^2(B)$.  For the pure $Q_3$ state at $h\to0^+$ we find $K=0.5855553284\ldots$, so the leading transverse exponent is
\begin{equation}
  \frac{1}{2K}=0.85389027\ldots .
  \label{eq:Cpm-exponent}
\end{equation}
The selected state is therefore a magnetized critical phase: it has a finite longitudinal spin density and finite chirality, but its connected transverse correlations decay algebraically.

We also checked Eq.~\eqref{eq:offsite-average} by exact diagonalization of the periodic pure-$Q_3$ selector.  For $L=6,8,10$, the zero-field ground multiplets have $S=1,2,2$, and the highest-weight sectors give global off-site averages $-0.066667$, $-0.035714$, and $-0.033333$, exactly matching Eq.~\eqref{eq:offsite-average}.  Representative longest-distance transverse correlators are $C^{+-}(L/2)=0.098568$, $-0.103357$, and $-0.078883$, respectively; the small sizes show strong oscillations, but no size-independent positive plateau.  This numerical check supports the analytic interpretation: the finite-volume state selected by $h\to0^+$ is a highest-weight representative of a macroscopic multiplet, whereas the thermodynamic phase is symmetry-broken because $S/L$ remains finite and a one-sided field selects a definite magnetization axis.  A globally rotated thermodynamic representative would have the corresponding rotated one-point magnetization. After subtracting that one-point piece, the connected transverse fluctuations remain algebraic.

\subsection{Counting of selected states.}
It is useful to separate the number of charge-selected magnetochiral states from the total number of states in the spin chain.  For a chain of length $L$, the full Hilbert space has dimension
\begin{equation}
  \dim{\cal H}_L=2^L ,
\end{equation}
and therefore carries the entropy density $\ln 2$.  The selector states discussed here form a much smaller set.  This counting refers to the extremal zero-temperature states exposed as ground states of the selector Hamiltonian, not to the much larger set of generic finite-entropy eigenstates that may have a nonzero value of $Q_3$.

For the pure current selector at $h\to0^+$ and fixed sign of $\alpha$, the finite-size state is the highest-weight representative of one macroscopic $SU(2)$ multiplet.  Its total spin scales as
\begin{equation}
  S_*(L)=m_* L+o(L),
  \qquad
  m_* \equiv M(0^+)=0.215636\ldots .
\end{equation}
The full multiplet therefore contains only
\begin{equation}
  2S_*(L)+1
  =
  2m_*L+o(L)
\end{equation}
orthogonal finite-volume states.  A longitudinal field $h\to0^+$ selects the highest-weight state
$\lvert S_*,S_*\rangle$, while $h\to0^-$ selects the lowest-weight state
$\lvert S_*,-S_*\rangle$.  Thus, for a fixed chirality orientation and a fixed symmetry-breaking axis, the two one-sided limits select two extremal representatives; allowing all members of the finite-volume multiplet still gives only an $O(L)$ number of orthogonal states.  A different infinitesimal field direction selects a different thermodynamic orientation, but these orientations arise from the same polynomial-size multiplet rather than from an exponentially large manifold.

If the field $h$ is varied along the pure-$Q_3$ magnetochiral branch, the total spin changes in discrete finite-size steps.  The highest-weight sectors then form an $O(L)$ set, with
\begin{equation}
  {\cal N}^{(Q_3)}_{\rm hw}(L)
  \sim
  \left(\frac12-m_*\right)L
\end{equation}
for one chirality orientation.  If all $SU(2)$ descendants of these multiplets are included, the count becomes
\begin{equation}
  {\cal N}^{(Q_3)}_{\rm mult}(L)
  \sim
  \sum_{S=S_*}^{L/2}(2S+1)
  =
  \left(\frac14-m_*^2\right)L^2+o(L^2).
\end{equation}
Including both signs of the chirality only multiplies these estimates by two.

More generally, in the finite-interval one-string branch selected by varying the ratios $J/\alpha$ and $h/\alpha$, a finite-size zero-entropy Bethe sea is specified by its two Fermi quantum numbers.  The number of such sharp intervals is therefore at most $O(L^2)$, and including their $SU(2)$ descendants gives at most a polynomial number of states, $O(L^3)$.  In all cases,
\begin{equation}
  \lim_{L\to\infty}\frac{1}{L}
  \log {\cal N}_{\rm sel}(L)=0,
  \qquad
  \lim_{L\to\infty}\frac{1}{L}
  \log \dim{\cal H}_L=\ln2 .
\end{equation}
The charge-selected magnetochiral states are therefore exponentially rare in the full Hilbert space.  This polynomial counting is the finite-size counterpart of the vanishing Yang--Yang entropy density: the selector exposes extremal Bethe macrostates rather than the exponentially many finite-entropy configurations that dominate thermal and generalized Gibbs ensembles.

\subsection{Generalized equilibrium, robustness, and transport}
\label{subsec:ghd-main}

The ensemble selected by the construction is the zero-generalized-temperature limit
\begin{equation}
  \rho_{\rm sel}=\lim_{\beta\to\infty}
  \frac{\exp[-\beta(JQ_2+\alpha Q_3-hS^z)]}
       {\mathrm{Tr}\,\exp[-\beta(JQ_2+\alpha Q_3-hS^z)]} .
  \label{eq:selected-ensemble-main}
\end{equation}
Because $Q_2$, $Q_3$, and $S^z$ commute with $H_{\rm XXX}$, this selects an eigenstate sector of the original chain rather than a thermal mixture.  At the trace-center point in Eq.~\eqref{eq:center-params}, a Gibbs ensemble of $H_{\rm XXX}$ would be the infinite-temperature ensemble, with no chirality and no preferred spin direction.  The charge-selected state instead has finite chirality, finite magnetization, algebraic correlations, and zero Yang--Yang entropy density.  It is therefore an exactly solvable atypical Bethe macrostate outside the finite-entropy manifold that dominates GGEs.

At $h=0$, the selector $H_{\chi0}=\alpha Q_3$ is $SU(2)$ invariant.  A finite chain therefore has a spin multiplet rather than a state with a chosen axis.  The thermodynamic symmetry-broken states are obtained by taking $L\to\infty$ first and then $h\to0^\pm$, which selects the highest- or lowest-weight component.  This would be a trivial label choice if the multiplet had $S=O(1)$.  Here the TBA gives $S/L=M(0^+)=0.215636\ldots$, so the selected component has a finite local magnetization density.  The one-sided magnetization in Eq.~\eqref{eq:zero} is therefore the order parameter for a macroscopic spin-sector selection, while the scalar chirality diagnoses the simultaneous current-carrying character of the same Bethe macrostate.

\begin{figure}[t]
    \centering
    \includegraphics[width=\linewidth]{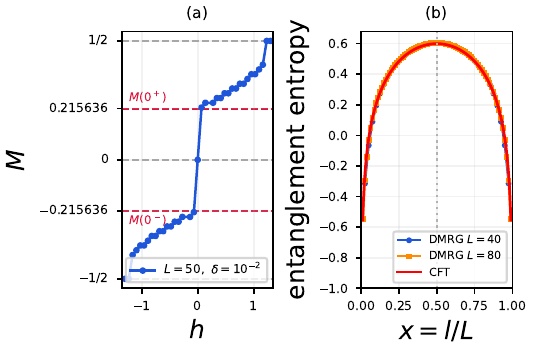}
    \caption{Finite-size benchmarks for the weakly nonintegrable Hamiltonian in Eq.~\eqref{eq:delta}.  (a) Magnetization density $M$ versus field for $L=50$ and $\delta=10^{-2}$.  The red dashed lines mark the exact pure-selector one-sided TBA values $M(0^\pm)=\pm0.215636\ldots$; the DMRG data remain close to this finite-magnetization scale near zero field and approach $|M|=1/2$ at large $|h|$.  (b) Bipartite entanglement entropy for $L=40$ and $L=80$, compared with the $c=1$ CFT form.  The figure was redrawn from the vector data in the supplied DMRG plots.  The persistence of a finite magnetization and a single-boson entanglement collapse supports a prethermal continuation of the integrable magnetochiral Luttinger liquid under weak integrability breaking.}
    \label{fig:dmrgdelta}
\end{figure}

We tested the stability of this picture by adding a weak nonintegrable Ising perturbation,
\begin{equation}
  H_{\chi h}^{(\delta)}=\alpha Q_3-hS^z+
  \delta\sum_j S_j^zS_{j+1}^z,
  \qquad \delta=10^{-2}.
  \label{eq:delta}
\end{equation}
This perturbation breaks the exact Bethe-ansatz hierarchy and explicitly fixes the spin axis, so it is not an independent test of spontaneous $SU(2)$ breaking.  It is instead a test of whether the magnetized, current-carrying critical sector remains visible once exact integrability is weakly relaxed.  In Figure~\ref{fig:dmrgdelta} we show that the one-sided magnetization remains finite near $h=0$, and that the entanglement data follow the $c=1$ Calabrese-Cardy form. Thus the weak perturbation rounds the ideal Bethe-ansatz structure only weakly on these system sizes: the state still looks like a magnetized, current-carrying Luttinger liquid before eventual long-time thermalization can set in.

As a complementary finite-size overlap and dynamical check, we compared the ground state of the integrable selector with the ground state after the same weak Ising perturbation.  We performed the direct overlap calculation by exact diagonalization on open chains, using the same local operators and $h=10^{-2}$ to select one spin sector.  The fidelity
\begin{equation}
  F_L(\delta)=|\langle\Psi_L(0)|\Psi_L(\delta)\rangle|^2
  \label{eq:fidelity-delta}
\end{equation}
remains very close to unity for $\delta=10^{-2}$: $F_6=0.999849$, $F_8=0.999990$, and $F_{10}=0.999910$.  We also computed the survival probability $|\langle\Psi_L(0)|e^{-iH_{\chi h}^{(\delta)}t}|\Psi_L(0)\rangle|^2$ for $L=10$; it stays above $0.99965$ up to $t=200$ in units of $\alpha^{-1}$.  These finite-system numbers show that the weak perturbation changes the selected state only perturbatively on accessible finite sizes and supports the prethermal interpretation.

In the integrable limit, long-wavelength dynamics is described by generalized hydrodynamics~\cite{castroalvaredo2016,bertini2016,alba2018}.  At zero generalized temperature the occupations are exactly $0$ or $1$, so the leading Euler-scale response is ballistic and is carried by dressed quasiparticles near the Fermi edges.  At finite generalized temperature, finite size, or weak integrability breaking, occupation noise reappears and eventually leads to diffusive or anomalous hydrodynamic behavior, including the spin-diffusive and superdiffusive regimes discussed for integrable and nearly integrable spin chains~\cite{Gopalakrishnan2018,Vasseur2018,bulchandani2021,singh2024,ulk2024,WangMoore2025}.  The sharp signatures of the present state are therefore the one-sided magnetization, the finite scalar chirality, the Luttinger-liquid correlations, and a prethermal window of ballistic spin and chirality transport.

Having identified these theoretical signatures, we now turn to experimental routes for realizing or approximating the current selector and measuring the resulting order parameters.

\section{Experimental realization of \texorpdfstring{$H_{\chi h}(\alpha,h)$}{H chi h(alpha,h)}} 
Beyond serving as a benchmark, engineering such charge-coupled Hamiltonians provides a preparation route for targeted high-energy sectors whenever the corresponding charges are conserved, or at least long lived, on experimental time scales.  A strictly one-dimensional Hubbard chain with only nearest-neighbor Peierls phases is not sufficient for a local scalar-chirality term, because the phases can be gauged to boundary twists.  The natural geometry is instead a zigzag chain or triangular ladder, where three neighboring sites form an elementary oriented plaquette.

A minimal microscopic model is a two-component Hubbard model on such triangles,
\begin{align}
H_{\rm Hub}=&-\sum_{\triangle ij,\sigma}
\bigl(t_{ij,\sigma}e^{i\phi_{ij,\sigma}}
 c_{i\sigma}^{\dagger}c_{j\sigma}+{\rm H.c.}\bigr)\nonumber\\
&+U\sum_i n_{i,\uparrow}n_{i,\downarrow}
-h\sum_i S_i^z ,
\label{eq:triangular-hubbard}
\end{align}
where the directed product of hopping phases around a plaquette gives a flux $\Phi_\sigma=\sum_{\triangle}\phi_{ij,\sigma}$.  Raman-assisted hopping and synthetic gauge fields allow the magnitudes and phases of the spin-resolved hoppings to be tuned \cite{JimenezGarcia2012Peierls,Goldman2014LightInducedGauge}.  In the Mott regime, $|t_{ij,\sigma}|\ll U$, a Schrieffer--Wolff expansion gives a spin model on the same zigzag backbone \cite{MacDonaldGirvinYoshioka1988}.  The second-order terms produce two-spin exchange on the bonds of the ladder.  For spin-independent hopping this is the usual Heisenberg exchange, $J_{ij}=4|t_{ij}|^2/U$.  Spin-dependent amplitudes or phases more generally generate anisotropic exchange, including $XXZ$-type couplings and Dzyaloshinskii--Moriya terms.  These terms need not destroy the desired regime, but they must be made small, compensated, or included as controlled perturbations to the ideal selector.

At third order, virtual hopping around an oriented triangle generates a ring-exchange contribution.  Its part odd under reversal of the plaquette orientation is the scalar chirality,
\begin{equation}
  K_\chi\sum_{\triangle ijk}
  {\bf S}_i\cdot({\bf S}_j\times{\bf S}_k),
  \qquad
  K_\chi \sim \frac{24 |t_{ij}t_{jk}t_{ki}|}{U^2}\sin\Phi,
  \label{eq:triangular-chirality}
\end{equation}
with geometry-dependent signs set by the orientation of the triangles.  This is the standard strong-coupling mechanism by which flux through triangular plaquettes produces chiral three-spin interactions \cite{motrunich2006,greiter2014,Pachos2004ThreeSpin}.  Scalar chirality also plays a central role as an order parameter and diagnostic for topologically ordered chiral spin liquids in frustrated-band and Chern--Simons descriptions \cite{SedrakyanGlazmanKamenev2014,SedrakyanGlazmanKamenev2015,MaitiSedrakyan2019,SedrakyanMoessnerKamenev2020,WangXieWangSedrakyan2022}.  Identifying \(\alpha\simeq K_\chi\), the spin Hamiltonian contains the same local operator \(Q_3\) that defines the magnetochiral selector.  The longitudinal field \(-h\sum_i S_i^z\) is simply the differential Zeeman or AC Stark shift between the two hyperfine states.

The practical goal is to make the chiral coupling large compared with unwanted two-spin terms.  Zigzag or triangular-ladder geometries help because the scalar chirality is generated on elementary plaquettes, while competing exchanges can be frustrated by geometry.  Floquet modulation of the lattice depth, hopping amplitudes, or interaction strength can further renormalize the second-order exchange relative to the third-order chiral term \cite{Chaudhary2019FloquetExchange}.  In an experiment one therefore aims for a hierarchy in which the dynamics over the observation window is dominated by \(K_\chi Q_3-hS^z\), while residual anisotropic exchanges produce only slow prethermal relaxation.

The resulting magnetochiral state can be prepared either by a quasi-adiabatic ramp of the synthetic flux and drive parameters, starting from a simple Mott product state, or by a quench into the chiral regime followed by prethermal relaxation.  Its properties are directly accessible with present quantum-gas-microscope technology \cite{Kuhr2016QGM}. Single-site, spin-resolved imaging of fermions has been demonstrated in one-dimensional and frustrated geometries, including triangular optical lattices \cite{Yang2021TriangularQGM}, enabling direct measurement of local magnetization profiles, spin--spin correlators, and, via measurements in several spin bases, the three-spin correlators entering the scalar chirality. These measurements allow experimental reconstruction of the order parameters \(M(\alpha,h)\) and \(\chi(\alpha,h)\), as well as the real-space current patterns associated with \(Q_3\).

The most direct observables are therefore: (i) the population imbalance $M=L^{-1}(N_\uparrow-N_\downarrow)/2$; (ii) the local scalar chirality $C_j=\langle\mathbf S_j\!\cdot(\mathbf S_{j+1}\times\mathbf S_{j+2})\rangle$, reconstructed from three-spin correlators in rotated spin bases; (iii) equal-time spin correlations, whose power-law decay diagnoses the Luttinger liquid; and (iv) dynamical current response or domain-wall spreading, which distinguishes the prethermal ballistic regime from late-time diffusion in weakly nonintegrable samples. Entanglement scaling is less direct experimentally, but it can be inferred in small cold-atom chains from randomized-measurement protocols and benchmarked numerically as in Figure~\ref{fig:dmrgdelta}.
Complementary realizations are also available in programmable Rydberg-atom and trapped-ion platforms, where multispin interactions are synthesized directly at the level of effective spin models. Proposals for engineering chiral and Dzyaloshinskii--Moriya--type interactions with Rydberg arrays \cite{Kuznetsova2023RydbergChiral,ValenciaTortora2024RydbergChiral} provide an alternative route to Hamiltonians closely related to $H_{\chi h}$, with independent tunability of $\alpha$ and $h$ and native access to out-of-equilibrium magnetochiral dynamics. Together, the analytic TBA solution, the weakly nonintegrable DMRG benchmarks, and these direct observables make the charge-selected magnetochiral state a concrete target for testing symmetry breaking in high-energy current-carrying sectors.

These measurement channels connect the exact selector construction to possible laboratory diagnostics.  We close by summarizing the physics that distinguishes the magnetochiral eigenstate from ordinary thermal high-energy states.

\section{Conclusions}

We have shown that conserved charges can be used as practical selectors of rare high-energy sectors in an interacting many-body spectrum.  Instead of searching directly through exponentially many excited states, one tilts the Hamiltonian by a conserved, or sufficiently long-lived, charge and studies the zero-temperature state of the tilted problem.  This turns an excited-state question for the original model into an equilibrium problem for the selector, while still identifying a definite eigenstate sector of the original Hamiltonian.  For the XXX chain this procedure gives exact atypical ordered, critical, zero-entropy Bethe macrostates at arbitrary XXX energy density, including the full Hilbert-space trace center.  These states are exact XXX eigenstates with finite energy density, finite scalar chirality, algebraic correlations, and local observables outside the finite-entropy GGE manifold.

For the spin-$\tfrac12$ XXX chain, the minimal selector $H_{\chi h}=\alpha Q_3-hS^z$ isolates a magnetochiral eigenstate whose properties are determined by a finite-interval TBA.  The finite-interval TBA gives the dressed energy, root density, free energy, magnetization, scalar chirality, saturation field, and the location of the state in the undeformed XXX spectrum from the same convention.  In the one-sided zero-field limit we find
\begin{align}
M(0^\pm)&=\pm0.215636\ldots,
&\chi(0)&=-0.245150\ldots,\nonumber\\
 e_{\rm XXX}^{\chi}&=-0.123967\ldots .
\end{align}
 At finite size the selected state is the highest- or lowest-weight representative of a giant $SU(2)$ multiplet. The thermodynamic order parameter is the nonzero spin density of that multiplet.  The energy-tuned selector in Eq.~\eqref{eq:HJchih-intro} reaches the trace-center energy density 
 while still retaining finite chirality and finite magnetization.  Varying $J/\alpha$ moves the selected eigenstate through the XXX many-body band: the magnetochiral branch connects the ferromagnetic edge to the antiferromagnetic edge, while the trace-center point is an ordered, zero-entropy representative at the unrestricted full-Hilbert-space infinite-temperature energy density.

The pure current selector reaches the trace center at finite field $h_*/h_c=0.62865083\ldots$, while the generalized selector reaches it at zero field by tuning $J_0/\alpha=-0.52207095\ldots$.  In both cases the state remains magnetochiral; the zero-field tuning has the additional advantage that spin-sector selection is spontaneous rather than explicitly imposed by a finite field.

The zero entropy density has a direct Bethe-ansatz meaning.  The selector fixes a sharp Bethe sea: the 1-string occupation is either filled or empty, and all higher strings are absent.  The Yang--Yang entropy integrand therefore vanishes point by point.  The state is not a thermal mixture of many locally distinct Bethe configurations.  It is a single zero-entropy macrostate.  Its criticality comes from a different source: particle-hole excitations near the Fermi edges cost arbitrarily little energy and generate a $c=1$ Luttinger liquid with logarithmic entanglement.

The weakly nonintegrable DMRG benchmarks show that the one-sided magnetization and the $c=1$ entanglement structure survive a small Ising perturbation on the available system sizes.  Exact integrability is required for the finite-interval TBA solution and for strictly ballistic Euler-scale dynamics, but the numerical results indicate a broader prethermal regime in which the magnetochiral critical sector remains visible.  This makes the state a realistic target for cold-atom, Rydberg, and trapped-ion platforms, where spin imbalance, scalar chirality, correlation exponents, and dynamical spreading can be measured directly.

Several extensions are natural.  Higher odd conserved charges can select more structured current-carrying Bethe seas and may reveal a larger family of symmetry-broken excited states.  A controlled theory of weak integrability breaking should determine the lifetime of the magnetochiral prethermal regime and its crossover to ordinary thermal transport.  Finally, experimental implementation would provide a direct test of high-energy order, zero-entropy criticality, and charge-selected dynamics in a setting where the relevant order parameters are local and measurable.

\begin{acknowledgments}
TAS thanks Nanjing University for its hospitality during a sabbatical visit. Part of the research reported
in this paper was performed at the Max-Planck-Institut für Physik komplexer Systeme, whose support is gratefully acknowledged. This work has been supported by the Armenian Higher Education and Science Committee under the ARPI Remote Laboratory program grant 24RL-1C024, research projects 21AG-1C024 and 25PostDoc1C003, and by the National R$\&$D Program of China grant 2022YFA1403601 and the Natural Science Foundation of Jiangsu Province grant BK20233001.
\end{acknowledgments}

\appendix

\section{Finite-interval thermodynamic Bethe ansatz}
\label{app:tba}

In this appendix, we give the details behind the finite-interval TBA used in Sec.~\ref{sec:benchmark-corrected}.  The key point is that the selected state is a finite Fermi sea of Bethe rapidities.  Consequently the energy, momentum, and charge response are dressed by the same finite-interval Bethe kernel.

\subsection{One-magnon derivation of the bare charges}
\label{app:one-magnon-charges}

In this subsection, we derive the one-magnon charge eigenvalues quoted in Eq.~\eqref{eq:magnon-dispersion}.  We write $\ket{j}=S_j^-\ket{F}$ for a localized spin flip.  Since $Q_2\ket{F}=(L/4)\ket{F}$, the one-magnon energy entering the selector is the eigenvalue of $Q_2-L/4$ in this sector.  The two bonds adjacent to the flipped spin give
\begin{equation}
  \left(Q_2-\frac{L}{4}\right)\ket{j}
  =\frac12\ket{j+1}+\frac12\ket{j-1}-\ket{j} .
\end{equation}
Acting on the plane wave $\ket{k}=L^{-1/2}\sum_j e^{\ii kj}\ket{j}$ therefore gives
\begin{align}
  \left(Q_2-\frac{L}{4}\right)\ket{k}
  &=\left(\frac12e^{\ii k}+\frac12e^{-\ii k}-1\right)\ket{k}
  \nonumber\\
  &=\left(\cos k-1\right)\ket{k} .
\end{align}
Thus $q_2(k)=\cos k-1$.

For the chiral charge we use the convention
\begin{equation}
  Q_3=-\sum_\ell {\cal X}_\ell,
  \qquad
  {\cal X}_\ell=\mathbf S_\ell\cdot
  \left(\mathbf S_{\ell+1}\times\mathbf S_{\ell+2}\right) .
\end{equation}
Only three chirality densities contain the flipped spin.  Their combined action gives
\begin{equation}
  Q_3\ket{j}
  =\frac{\ii}{4}\left(\ket{j+2}-\ket{j-2}\right)
   -\frac{\ii}{2}\left(\ket{j+1}-\ket{j-1}\right) .
\end{equation}
Consequently,
\begin{align}
  Q_3\ket{k}
  &=\left[\frac{\ii}{4}\left(e^{-2\ii k}-e^{2\ii k}\right)
  -\frac{\ii}{2}\left(e^{-\ii k}-e^{\ii k}\right)\right]\ket{k}
  \nonumber\\
  &=\left(\frac12\sin2k-\sin k\right)\ket{k}
  =-\sin k\,(1-\cos k)\ket{k} .
\end{align}
This proves the expression for $q_3(k)$ in Eq.~\eqref{eq:magnon-dispersion}.  Using Eq.~\eqref{eq:rapidity-map} then gives the rapidity-space charges in Eq.~\eqref{eq:one-string-bare-charges}.

\subsection{Bare charges and dressed energy}

In the rapidity convention used in the main text, the XXX kernels are those in Eq.~\eqref{eq:akernel}.  The bare one-string charges needed here are
\begin{align}
  q_2^{(1)}(x)&=-2\pi a_1(x)=-\frac{2}{x^2+1},\nonumber\\
  q_3^{(1)}(x)&=2\pi a_1'(x)=-\frac{4x}{(x^2+1)^2} .
\end{align}
The sign of $q_3^{(1)}$ fixes the orientation of the chiral sector.  Reversing the sign of $\alpha$ selects the parity- and time-reversed sector.

For $h>0$ and $0<h<h_c$, the occupied rapidities form one interval $\mathcal I=[x_-,x_+]$.  The finite-interval dressed energy is Eq.~\eqref{eq:eps1}; the endpoints are fixed by Eq.~\eqref{eq:Fpoints}.  This is the zero-temperature form of the Bethe-Takahashi variational problem restricted to the one-string phase.  For the pure current selector, $J=0$, the same equations also make the saturation field transparent.  The interval closes when the minimum of $d_{0,h}(x)=h+\alpha q_3^{(1)}(x)$ first touches zero.  Since the minimum occurs at $x_0=1/\sqrt3$, one obtains Eq.~\eqref{eq:hc-corrected}.

\subsection{Density equation and free energy}

The density of occupied roots is dressed on the same interval and obeys Eq.~\eqref{eq:rho-corrected-secIII}.  The free energy can be written in the density form Eq.~\eqref{eq:f-density}.  To prove the equivalent form Eq.~\eqref{eq:f}, multiply Eq.~\eqref{eq:eps1} by $\rho(x)$ and Eq.~\eqref{eq:rho-corrected-secIII} by $\varepsilon(x)$, integrate over $\mathcal I$, and subtract.  Because $a_2(x-y)=a_2(y-x)$, the double integrals cancel and one obtains
\begin{equation}
  \int_{\mathcal I}\!dx\,\rho(x)d_{J,h}(x)
  =\int_{\mathcal I}\!dx\,a_1(x)\varepsilon(x).
\end{equation}
This identity is useful in numerics because it checks that the dressed energy and density have been solved in the same convention.

The order parameters follow by differentiating the free energy.  For example, changing $h$ changes both the driving term and the endpoints.  The endpoint contributions vanish because $\varepsilon(x_\pm)=0$, leaving Eq.~\eqref{eq:M}.  The same argument for $\alpha$ gives Eq.~\eqref{eq:chi}.  The XXX energy density is obtained by inserting the bare $Q_2$ charge into the same root density, giving Eq.~\eqref{eq:xxx-energy-hzero} for the pure $Q_3$ selector and the trace-center condition quoted in Eq.~\eqref{eq:center-params} for the energy-tuned selector.

\subsection{Higher-string stability}
\label{app:higher-string-stability}

In the TBA in the main text, we used a one-string finite interval.  A self-consistency check is that no higher Bethe string has negative dressed energy in this background.  The bare charge carried by an $n$-string is
\begin{equation}
  q_2^{(n)}(x)=-\frac{2n}{x^2+n^2},
  \qquad
  q_3^{(n)}(x)=-\frac{4nx}{(x^2+n^2)^2},
  \label{eq:bare-n-string-charges}
\end{equation}
and its bare driving term is
\begin{equation}
  d_n(x)=nh+Jq_2^{(n)}(x)+\alpha q_3^{(n)}(x) .
\end{equation}
After the $1$-string sea is filled, the energy cost of inserting a dilute $n\ge2$ string is dressed by its scattering with that sea.  The corresponding test pseudoenergy is
\begin{align}
  \varepsilon_n^{\rm test}(x)
  &=d_n(x)-\int_{\mathcal I}dy\,
  K_{n1}(x-y)\varepsilon(y),
  \qquad n\ge2,
  \nonumber\\
  K_{n1}(x)&=a_{n-1}(x)+a_{n+1}(x).
  \label{eq:higher-string-test}
\end{align}
The one-string phase is stable when $\varepsilon_n^{\rm test}(x)>0$ for all finite $x$ and all $n\ge2$.  We evaluated Eq.~\eqref{eq:higher-string-test} with the same finite-interval TBA code used for the figures.  Throughout the magnetochiral branch no higher string acquires negative dressed energy.  At strictly $h=0^+$ the minimum approaches zero only as $|x|\to\infty$, which is the global spin-rotation zero mode of the $SU(2)$ multiplet rather than a finite-rapidity instability toward a thermodynamic density of higher strings.

\subsection{Exact zero Yang--Yang entropy}

 The zero-temperature selector fills a sharp interval of $1$-string rapidities and leaves all higher strings empty.  In terms of the occupation functions introduced in Eq.~\eqref{eq:tba_occ_main},
\begin{equation}
  n_1(x)=\Theta[-\varepsilon(x)],
  \qquad n_m(x)=0\quad(m\ge2).
\end{equation}
The Yang--Yang entropy density is given by Eq.~\eqref{eq:YY_main}.  Since the local binary entropy satisfies ${\cal S}(0)={\cal S}(1)=0$, every point of the rapidity axis gives zero contribution.  The Fermi edges themselves have measure zero.  Therefore
\begin{equation}
  s_{\rm YY}=0
\end{equation}
exactly for the whole finite-interval branch, including the $h\to0^+$ magnetochiral state and the energy-tuned trace-center state.  This zero-entropy property is the simplest analytic result of the TBA solution.  It explains why the selected state can lie deep in the XXX spectrum, even at the trace center, without becoming an infinite-temperature thermal state.
\subsection{Numerical solution}

The numerical values in Sec.~\ref{sec:benchmark-corrected} were obtained by a Nystr\"om discretization with Gauss--Legendre quadrature.  For a trial interval $[x_-,x_+]$, the integral equation for $\varepsilon$ becomes a dense linear system.  The two nonlinear equations $\varepsilon(x_-)=\varepsilon(x_+)=0$ are then solved by Newton iteration or a hybrid root finder.  Once the interval is fixed, the density equation is another linear system on the same quadrature nodes.  This procedure gives, for $\alpha=1$,
\begin{align}
  h_c&=1.299038105676\ldots,\nonumber\\
  M(0^+)&=0.215636088265\ldots,\nonumber\\
  \chi(0)&=-0.245150231693\ldots,\nonumber\\
  e_{\rm XXX}^{\chi}&=-0.123966864540\ldots.
\end{align}
The zero-field trace-center point obtained by tuning $J/\alpha$ is
\begin{align}
  J_0/\alpha&=-0.522070946406\ldots,\nonumber\\
  M_0&=0.281215249834\ldots,\nonumber\\
  \chi_0&=-0.211881874066\ldots .
\end{align}
We used Mathematica to reproduce these values, data and Figs.~\ref{fig:pureQ3profile}--\ref{fig:global-phase}. The same numerical solution also gives the trace-center point of the pure current selector.  Solving Eq.~\eqref{eq:hstar-definition} gives
\begin{align}
  h_*/h_c&=0.628650832126\ldots,\nonumber\\
  M_*&=0.326869110235\ldots,\nonumber\\
  \chi_*&=-0.198225673818\ldots .
\end{align}
These numbers are quoted in the main text in Eq.~\eqref{eq:hstar-params}.  They provide a useful distinction between two ways of reaching the trace center.  A finite field reaches $e_{\rm XXX}=0$ within the pure $Q_3$ selector but explicitly chooses the spin sector.  The $JQ_2$ tuning in Eq.~\eqref{eq:center-params} reaches the same trace-center energy at $h\to0^+$, where the spin-sector selection is spontaneous.

The finite-interval solution fixes the static thermodynamics.  To obtain the critical theory around that state, we next dress the momentum, velocity, and charge response on the same interval.

\section{Dressed velocity and Luttinger-liquid data}
\label{sec:low-energy-cft-data}

The low-energy theory is obtained by expanding around the two Fermi edges of the same finite interval.  The derivative of the dressed momentum is
\begin{equation}
  p'(x)=2\pi\rho(x),
\end{equation}
where $\rho(x)$ solves Eq.~\eqref{eq:rho-corrected-secIII}.  Differentiating Eq.~\eqref{eq:eps1} gives
\begin{equation}
  \partial_x\varepsilon(x)
  +\int_{\mathcal I}\!dy\,a_2'(x-y)\varepsilon(y)
  =\partial_x d_{J,h}(x).
\end{equation}
For the pure current selector, $\partial_x d_{0,h}=\alpha\partial_xq_3^{(1)}$.  The branch velocities are the dressed ratios in Eq.~\eqref{eq:dressedvelocitymain}.  These are the velocities entering finite-size spectra and real-time correlation functions.

The dressed charge is computed from Eq.~\eqref{eq:dressedcharge-main}.  Since the kernel depends only on rapidity differences, shifting the interval to $[-B,B]$ gives a single endpoint value ${\cal Z}(B)$ and $K={\cal Z}^2(B)$, Eq.~\eqref{eq:K-main}.  Together, $K$ and $v_\pm$ determine the universal low-energy theory: $K$ fixes equal-time exponents, while $v_\pm$ fix the two chiral propagation speeds.  The finite-size and low-generalized-temperature corrections quoted in Eqs.~\eqref{eq:finite-size-cft} and~\eqref{eq:lowT-main} follow from one compact boson with generally unequal branch velocities.

\subsection{Low-field expansion}

For the pure current selector at $h\to0^+$, the interval approaches the finite interval $\mathcal I_0=[x_-^0,x_+^0]$ quoted in Eq.~\eqref{eq:hzero-endpoints-corrected}.  Because the Fermi edges remain finite, the small-field expansion is regular.  Write
\begin{equation}
  \varepsilon(x;h)=\varepsilon_0(x)+h\eta(x)+O(h^2),
\end{equation}
with all functions in this subsection evaluated on $\mathcal I_0$.  Linearizing Eq.~\eqref{eq:eps1} gives the dressed response
\begin{equation}
  \eta(x)+\int_{\mathcal I_0}\!dy\,a_2(x-y)\eta(y)=1.
  \label{eq:eta-linear}
\end{equation}
The Fermi edges move according to
\begin{equation}
  \frac{d x_a}{dh}\bigg|_{0^+}
  =-\frac{\eta(x_a^0)}{\varepsilon_0'(x_a^0)},
  \qquad a=\pm .
  \label{eq:edge-linear}
\end{equation}
Equations~\eqref{eq:eta-linear} and~\eqref{eq:edge-linear}, together with the linearized density and dressed-charge equations, give the low-field expansion of $v_\pm$ and $K$.  Numerically, for the normalization $\alpha=1$ used in the figures,
\begin{align}
  \frac{v_+}{h_c}&=0.4610266712
  +17.46427417\frac{h}{h_c}+O[(h/h_c)^2],\nonumber\\
  \frac{v_-}{h_c}&=-1.711698524
  +0.214980962\frac{h}{h_c}+O[(h/h_c)^2],\nonumber\\
  K&=0.5855553284
  +0.998991823\frac{h}{h_c}+O[(h/h_c)^2].
  \label{eq:lowfield-appendix}
\end{align}
This is Eq.~\eqref{eq:lowfield-vK-main} in the main text.  For $h\to0^-$, spin reversal maps the interval to its mirror image and exchanges the two branches.  Thus the equal-time parameter $K$ is a function of $|h|$, while the signed velocities are branch-exchanged.

\subsection{Saturation expansion}

Near saturation, $h\to h_c^-$, the interval is a small pocket around the one-magnon minimum $x_0=1/\sqrt3$.  Let
\begin{equation}
  \delta=1-h/h_c,
  \qquad x=x_0+y.
\end{equation}
For $J=0$ and $\alpha=1$,
\begin{equation}
  q_3^{(1)}(x_0)=-h_c,
  \qquad
  \partial_x^2q_3^{(1)}(x_0)=\frac{27\sqrt3}{8}.
\end{equation}
The leading driving term is therefore
\begin{equation}
  d_{0,h}(x)= -h_c\delta
  +\frac12\partial_x^2q_3^{(1)}(x_0)y^2+\cdots .
\end{equation}
The pocket half-width is
\begin{equation}
  B_{\rm sat}=\frac{2}{3}\sqrt{\delta}+O(\delta).
\end{equation}
Since the pocket is small, dressing by the kernel gives only subleading corrections at leading order.  The density is $\rho(x_0)=a_1(x_0)+O(B_{\rm sat})=3/(4\pi)+O(B_{\rm sat})$.  This immediately yields the square-root order-parameter laws in Eq.~\eqref{eq:saturationcheck}.

The same expansion gives the saturation velocities.  At the two edges,
\begin{equation}
  \varepsilon'(x_\pm)=\pm\frac{9\sqrt3}{4}\sqrt{\delta}+O(\delta),
  \qquad
  2\pi\rho(x_\pm)=\frac32+O(\sqrt\delta).
\end{equation}
Dividing by $h_c=3\sqrt3/4$ gives
\begin{equation}
  \frac{v_\pm}{h_c}=\pm2\sqrt{1-h/h_c}+O(1-h/h_c).
\end{equation}
For the Luttinger parameter, the dressed charge on a short interval satisfies
\begin{align}
  {\cal Z}(B_{\rm sat})&=1-
  \int_0^{2B_{\rm sat}}\!du\,a_2(u)+O(B_{\rm sat}^2)\nonumber\\
  &=1-\frac{B_{\rm sat}}{\pi}+O(B_{\rm sat}^2).
\end{align}
Therefore
\begin{equation}
  K=1-\frac{4}{3\pi}\sqrt{1-h/h_c}+O(1-h/h_c),
\end{equation}
which is Eq.~\eqref{eq:saturation-vK-main}.  The same formulas hold as functions of $|h|$ on the spin-reversed branch.

These expansions give analytic control at both ends of the field interval and provide consistency checks on the full finite-interval numerics.  They also make clear that the $h\to0^{\pm}$ values are one-sided thermodynamic limits, while the saturation edge is an ordinary dilute-pocket limit.

\bibliographystyle{apsrev4-2}
\bibliography{PRB-version2}

@CONTROL{apsrev42Control,
  title = {1}
}

@article{deutsch1991,
	author = {Deutsch, J. M.},
	doi = {10.1103/PhysRevA.43.2046},
	journal = {Phys. Rev. A},
	pages = {2046--2049},
	title = {{Quantum statistical mechanics in a closed system}},
	volume = {43},
	year = {1991}}

@article{srednicki1994,
	author = {Srednicki, M.},
	doi = {10.1103/PhysRevE.50.888},
	journal = {Phys. Rev. E},
	pages = {888--901},
	title = {{Chaos and quantum thermalization}},
	volume = {50},
	year = {1994}}

@article{rigol2008,
	author = {Rigol, M. and Dunjko, V. and Olshanii, M.},
	doi = {10.1038/nature06838},
	journal = {Nature},
	number = {7189},
	pages = {854--858},
	title = {{Thermalization and its mechanism for generic isolated quantum systems}},
	volume = {452},
	year = {2008}}

@article{dalessio2016,
	author = {D'Alessio, L. and Kafri, Y. and Polkovnikov, A. and Rigol, M.},
	doi = {10.1080/00018732.2016.1198134},
	journal = {Adv. Phys.},
	pages = {239--362},
	title = {{From quantum chaos and eigenstate thermalization to statistical mechanics and thermodynamics}},
	volume = {65},
	year = {2016}}

@article{PatilRigol2025PRB,
  author        = {Patil, Rohit and Rigol, Marcos},
  title         = {{Eigenstate thermalization in spin-$\frac{1}{2}$ systems with {SU}(2) symmetry}},
  journal       = {Phys. Rev. B},
  volume        = {111},
  pages         = {205126},
  year          = {2025},
  doi           = {10.1103/PhysRevB.111.205126},
  url           = {https://doi.org/10.1103/PhysRevB.111.205126},
  eprint        = {2503.01846},
  archivePrefix = {arXiv},
  primaryClass  = {quant-ph}
}

@article{bernien2017,
	author = {Bernien, Hannes and Schwartz, Sylvain and Keesling, Alexander and Levine, Harry and Omran, Ahmed and Pichler, Hannes and Choi, Soonwon and Zibrov, Alexander S. and Endres, Manuel and Greiner, Markus and Vuleti{\'c}, Vladan and Lukin, Mikhail D.},
	doi = {10.1038/nature24622},
	journal = {Nature},
	number = {7682},
	pages = {579--584},
	title = {{Probing many-body dynamics on a 51-atom quantum simulator}},
	volume = {551},
	year = {2017}}

@article{turner2018,
	author = {Turner, C. J. and Michailidis, A. A. and Abanin, D. A. and Serbyn, M. and Papi{\'c}, Z.},
	doi = {10.1038/s41567-018-0137-5},
	journal = {Nat. Phys.},
	pages = {745--749},
	title = {{Weak ergodicity breaking from quantum many-body scars}},
	volume = {14},
	year = {2018}}

@article{serbyn2021,
	author = {Serbyn, M. and Abanin, D. A. and Papi{\'c}, Z.},
	doi = {10.1038/s41567-021-01230-2},
	journal = {Nat. Phys.},
	pages = {675--685},
	title = {{Quantum many-body scars and weak breaking of ergodicity}},
	volume = {17},
	year = {2021}}

@article{nandkishore2015,
	author = {Nandkishore, R. and Huse, D. A.},
	doi = {10.1146/annurev-conmatphys-031214-014726},
	journal = {Annu. Rev. Condens. Matter Phys.},
	pages = {15--38},
	title = {{Many-body localization and thermalization in quantum statistical mechanics}},
	volume = {6},
	year = {2015}}

@article{abanin2019MBL,
	author = {Abanin, D. A. and Altman, E. and Bloch, I. and Serbyn, M.},
	doi = {10.1103/RevModPhys.91.021001},
	journal = {Rev. Mod. Phys.},
	pages = {021001},
	title = {{Colloquium: Many-body localization, thermalization, and entanglement}},
	volume = {91},
	year = {2019}}

@article{altman2015,
  author = {Altman, Ehud and Vosk, Ronen},
  title = {{Universal dynamics and renormalization in many-body-localized systems}},
  journal = {Annu. Rev. Condens. Matter Phys.},
  volume = {6},
  pages = {383--409},
  year = {2015},
  doi = {10.1146/annurev-conmatphys-031214-014701},
  url = {https://doi.org/10.1146/annurev-conmatphys-031214-014701}
}

@book{takahashi_book,
	address = {Cambridge},
	author = {Takahashi, M.},
	publisher = {Cambridge University Press},
	title = {{Thermodynamics of One-Dimensional Solvable Models}},
	year = {1999}}

@book{korepin_book,
	address = {Cambridge},
	author = {Korepin, V. E. and Bogoliubov, N. M. and Izergin, A. G.},
	publisher = {Cambridge University Press},
	title = {{Quantum Inverse Scattering Method and Correlation Functions}},
	year = {1993}}

@article{Berges2008,
  author = {Berges, J{\"u}rgen and Rothkopf, Alexander and Schmidt, Jonas},
  title = {{Nonthermal Fixed Points: Effective Weak Coupling for Strongly Correlated Systems Far from Equilibrium}},
  journal = {Phys. Rev. Lett.},
  volume = {101},
  pages = {041603},
  year = {2008},
  doi = {10.1103/PhysRevLett.101.041603}
}

@article{abanin2015prethermal,
	author = {Abanin, D. A. and De Roeck, W. and Ho, W. W. and Huveneers, F.},
	doi = {10.1007/s00220-017-2930-x},
	journal = {Commun. Math. Phys.},
	pages = {809--827},
	title = {{A rigorous theory of many-body prethermalization for periodically driven and closed quantum systems}},
	volume = {354},
	year = {2017}}

@article{abanin2017prethermal,
	author = {Abanin, D. A. and De Roeck, W. and Huveneers, F. and Ho, W. W.},
	doi = {10.1103/PhysRevB.95.014112},
	journal = {Phys. Rev. B},
	pages = {014112},
	title = {{Effective Hamiltonians, prethermalization, and slow energy absorption in periodically driven many-body systems}},
	volume = {95},
	year = {2017}}

@article{mori2018,
	author = {Mori, T.},
	doi = {10.1088/1361-6455/aabac0},
	journal = {J. Phys. B},
	pages = {112001},
	title = {{Thermalization and prethermalization in isolated quantum systems: a theoretical overview}},
	volume = {51},
	year = {2018}}

@article{bargheer2009,
  author = {Bargheer, Till and Beisert, Niklas and Loebbert, Florian},
  title = {{Long-Range Deformations for Integrable Spin Chains}},
  journal = {J. Phys. A: Math. Theor.},
  volume = {42},
  pages = {285205},
  year = {2009},
  doi = {10.1088/1751-8113/42/28/285205}
}

@article{smirnov2017ttbar,
  author = {Smirnov, Fedor A. and Zamolodchikov, Alexander B.},
  title = {{On space of integrable quantum field theories}},
  journal = {Nucl. Phys. B},
  volume = {915},
  pages = {363--383},
  year = {2017},
  doi = {10.1016/j.nuclphysb.2016.12.014}
}

@article{longrangeTTbar,
  author = {Pozsgay, Bal{\'a}zs and Jiang, Yunfeng and Tak{\'a}cs, G{\'a}bor},
  title = {{$T\bar{T}$-deformation and long range spin chains}},
  journal = {J. High Energy Phys.},
  volume = {2020},
  number = {3},
  pages = {092},
  year = {2020},
  doi = {10.1007/JHEP03(2020)092}
}

@article{cecile2024,
  title = {{Squeezed Ensembles and Anomalous Dynamic Roughening in Interacting Integrable Chains}},
  author = {Cecile, Guillaume and De Nardis, Jacopo and Ilievski, Enej},
  journal = {Phys. Rev. Lett.},
  volume = {132},
  issue = {13},
  pages = {130401},
  numpages = {6},
  year = {2024},
  month = {Mar},
  publisher = {American Physical Society},
  doi = {10.1103/PhysRevLett.132.130401},
  url = {https://link.aps.org/doi/10.1103/PhysRevLett.132.130401}
}

@article{krajnik2024,
  title = {{Dynamical Criticality of Magnetization Transfer in Integrable Spin Chains}},
  author = {Krajnik, {\v{Z}}iga and Schmidt, Johannes and Ilievski, Enej and Prosen, Toma{\v{z}}},
  journal = {Phys. Rev. Lett.},
  volume = {132},
  issue = {1},
  pages = {017101},
  numpages = {7},
  year = {2024},
  month = {Jan},
  publisher = {American Physical Society},
  doi = {10.1103/PhysRevLett.132.017101},
  url = {https://link.aps.org/doi/10.1103/PhysRevLett.132.017101}
}

@article{WMS2024,
  author = {Wei, Chenan and Mkhitaryan, Vagharsh V. and Sedrakyan, Tigran A.},
  title = {{Unveiling chiral states in the {XXZ} chain: finite-size scaling probing symmetry-enriched {$c=1$} conformal field theories}},
  journal = {J. High Energy Phys.},
  volume = {2024},
  number = {6},
  pages = {125},
  year = {2024},
  doi = {10.1007/JHEP06(2024)125},
  eprint = {2312.16660},
  archivePrefix = {arXiv},
  primaryClass = {cond-mat.str-el}
}

@article{rigol2007,
	author = {Rigol, M. and Dunjko, V. and Yurovsky, V. and Olshanii, M.},
	doi = {10.1103/PhysRevLett.98.050405},
	journal = {Phys. Rev. Lett.},
	pages = {050405},
	title = {{Relaxation in a Completely Integrable Many-Body Quantum System: An {Ab Initio} Study of the Dynamics of the Highly Excited States of 1D Lattice Hard-Core Bosons}},
	volume = {98},
	year = {2007}}

@article{CassidyClarkRigol2011,
  author  = {Cassidy, Amy C. and Clark, Charles W. and Rigol, Marcos},
  title   = {{Generalized thermalization in an integrable lattice system}},
  journal = {Phys. Rev. Lett.},
  volume  = {106},
  number  = {14},
  pages   = {140405},
  year    = {2011},
  doi     = {10.1103/PhysRevLett.106.140405}
}

@article{vidmar2016,
  author = {Vidmar, Lev and Rigol, Marcos},
  title = {{Generalized {Gibbs} ensemble in integrable lattice models}},
  journal = {J. Stat. Mech.},
  volume = {2016},
  pages = {064007},
  year = {2016},
  doi = {10.1088/1742-5468/2016/06/064007}
}

@article{Yuzbashyan2016,
  author  = {Yuzbashyan, Emil A.},
  title   = {{Generalized microcanonical and {G}ibbs ensembles in classical and quantum integrable dynamics}},
  journal = {Annals of Physics},
  volume  = {367},
  pages   = {288--296},
  year    = {2016},
  doi     = {10.1016/j.aop.2016.02.002}
}

@article{Reiter2021,
  author  = {Reiter, Florentin and Lange, Florian and Jain, Shreyans and Grau, Matt and Home, Jonathan P. and Lenar{\v{c}}i{\v{c}}, Zala},
  title   = {{Engineering generalized {G}ibbs ensembles with trapped ions}},
  journal = {Phys. Rev. Research},
  volume  = {3},
  pages   = {033142},
  year    = {2021},
  doi     = {10.1103/PhysRevResearch.3.033142}
}

@article{SedrakyanBabujian2022,
  author  = {Sedrakyan, Tigran A. and Babujian, Hrachya M.},
  title   = {{Quantum nonequilibrium dynamics from {Knizhnik-Zamolodchikov} equations}},
  journal = {J. High Energy Phys.},
  volume  = {2022},
  number  = {04},
  pages   = {039},
  year    = {2022},
  doi     = {10.1007/JHEP04(2022)039},
  url     = {https://doi.org/10.1007/JHEP04(2022)039},
  eprint  = {2112.12866},
  archivePrefix = {arXiv},
  primaryClass = {quant-ph}
}

@article{castroalvaredo2016,
  author = {Castro-Alvaredo, Olalla A. and Doyon, Benjamin and Yoshimura, Takato},
  title = {{Emergent Hydrodynamics in Integrable Quantum Systems Out of Equilibrium}},
  journal = {Phys. Rev. X},
  volume = {6},
  number = {4},
  pages = {041065},
  year = {2016},
  doi = {10.1103/PhysRevX.6.041065}
}

@article{bertini2016,
  author = {Bertini, Bruno and Collura, Mario and De Nardis, Jacopo and Fagotti, Maurizio},
  title = {{Transport in out-of-equilibrium XXZ chains: Exact profiles of charges and currents}},
  journal = {Phys. Rev. Lett.},
  volume = {117},
  number = {20},
  pages = {207201},
  year = {2016},
  doi = {10.1103/PhysRevLett.117.207201}
}

@article{CalabreseCardy2004,
  author       = {Calabrese, Pasquale and Cardy, John},
  title        = {{Entanglement entropy and quantum field theory}},
  journal      = {Journal of Statistical Mechanics: Theory and Experiment},
  volume       = {2004},
  number       = {06},
  pages        = {P06002},
  year         = {2004},
  doi          = {10.1088/1742-5468/2004/06/P06002},
  archivePrefix = {arXiv},
  eprint       = {hep-th/0405152}
}

@article{ilievski2016,
  author = {Ilievski, Enej and De Nardis, Jacopo and Wouters, Bram and Caux, Jean-S{\'e}bastien and Essler, Fabian H. L. and Prosen, Toma{\v z}},
  title = {{Complete generalized {Gibbs} ensembles in an interacting theory}},
  journal = {Phys. Rev. Lett.},
  volume = {115},
  number = {15},
  pages = {157201},
  year = {2015},
  doi = {10.1103/PhysRevLett.115.157201}
}

@article{odea2025,
  title = {{Entanglement Oscillations from Many-Body Quantum Scars}},
  author = {O'Dea, Nicholas and Sriram, Adithya},
  journal = {Phys. Rev. Lett.},
  volume = {134},
  issue = {21},
  pages = {210402},
  numpages = {9},
  year = {2025},
  month = {May},
  publisher = {American Physical Society},
  doi = {10.1103/PhysRevLett.134.210402},
  url = {https://link.aps.org/doi/10.1103/PhysRevLett.134.210402}
}

@article{alba2018,
  title = {{Entanglement and quantum transport in integrable systems}},
  author = {Alba, Vincenzo},
  journal = {Phys. Rev. B},
  volume = {97},
  issue = {24},
  pages = {245135},
  numpages = {12},
  year = {2018},
  month = {Jun},
  publisher = {American Physical Society},
  doi = {10.1103/PhysRevB.97.245135},
  url = {https://link.aps.org/doi/10.1103/PhysRevB.97.245135}
}

@article{Gopalakrishnan2018,
  author       = {Gopalakrishnan, Sarang and Vasseur, Romain},
  title        = {{Kinetic theory of spin diffusion and superdiffusion in {XXZ} spin chains}},
  journal      = {Phys. Rev. Lett.},
  volume       = {122},
  number       = {12},
  pages        = {127202},
  year         = {2019},
  doi          = {10.1103/PhysRevLett.122.127202}
}

@article{Vasseur2018,
  author       = {Gopalakrishnan, Sarang and Huse, David A. and Khemani, Vedika and Vasseur, Romain},
  title        = {{Hydrodynamics of operator spreading and quasiparticle diffusion in interacting integrable systems}},
  journal      = {Phys. Rev. B},
  volume       = {98},
  number       = {22},
  pages        = {220303(R)},
  year         = {2018},
  doi          = {10.1103/PhysRevB.98.220303}
}

@article{bulchandani2021,
  author = {Bulchandani, Vir B. and Gopalakrishnan, Sarang and Ilievski, Enej},
  title = {{Superdiffusion in spin chains}},
  journal = {J. Stat. Mech.},
  volume = {2021},
  number = {8},
  pages = {084001},
  year = {2021},
  doi = {10.1088/1742-5468/ac12c7}
}

@article{singh2024,
  title = {{Tunable Superdiffusion in Integrable Spin Chains Using Correlated Initial States}},
  author = {Singh, Hansveer and Kolodrubetz, Michael H. and Gopalakrishnan, Sarang and Vasseur, Romain},
  journal = {Phys. Rev. Lett.},
  volume = {132},
  issue = {17},
  pages = {176303},
  numpages = {6},
  year = {2024},
  month = {Apr},
  publisher = {American Physical Society},
  doi = {10.1103/PhysRevLett.132.176303},
  url = {https://link.aps.org/doi/10.1103/PhysRevLett.132.176303}
}

@article{ulk2024,
  title = {{Iterative Construction of Conserved Quantities in Dissipative Nearly Integrable Systems}},
  author = {Ul{\v{c}}akar, Iris and Lenar{\v{c}}i{\v{c}}, Zala},
  journal = {Phys. Rev. Lett.},
  volume = {132},
  issue = {23},
  pages = {230402},
  numpages = {7},
  year = {2024},
  month = {Jun},
  publisher = {American Physical Society},
  doi = {10.1103/PhysRevLett.132.230402},
  url = {https://link.aps.org/doi/10.1103/PhysRevLett.132.230402}
}

@article{WangMoore2025,
  author  = {Wang, Kevin and Moore, Joel E.},
  title   = {{Breakdown of superdiffusion in perturbed quantum integrable spin chains and ladders}},
  journal = {Phys. Rev. B},
  volume  = {112},
  pages   = {104434},
  year    = {2025},
  doi     = {10.1103/2hb5-k1jz},
  url     = {https://doi.org/10.1103/2hb5-k1jz},
  eprint  = {2501.08866},
  archivePrefix = {arXiv},
  primaryClass = {cond-mat.stat-mech}
}

@article{JimenezGarcia2012Peierls,
  title   = {{Peierls Substitution in an Engineered Lattice Potential}},
  author  = {Jim{\'e}nez-Garc{\'i}a, K. and LeBlanc, L. J. and Williams, R. A. and Beeler, M. C. and Perry, A. R. and Spielman, I. B.},
  journal = {Phys. Rev. Lett.},
  volume  = {108},
  pages   = {225303},
  year    = {2012},
  doi     = {10.1103/PhysRevLett.108.225303}
}

@article{Goldman2014LightInducedGauge,
  title   = {{Light-induced gauge fields for ultracold atoms}},
  author  = {Goldman, N. and Juzeli{\=u}nas, G. and {\"O}hberg, P. and Spielman, I. B.},
  journal = {Rep. Prog. Phys.},
  volume  = {77},
  pages   = {126401},
  year    = {2014},
  doi     = {10.1088/0034-4885/77/12/126401}
}

@article{motrunich2006,
  author = {Motrunich, Olexei I.},
  title = {{Orbital magnetic field effects in spin liquid with spinon Fermi sea: Possible application to $\kappa$-(ET)$_2$Cu$_2$(CN)$_3$}},
  journal = {Phys. Rev. B},
  volume = {73},
  pages = {155115},
  year = {2006},
  doi = {10.1103/PhysRevB.73.155115}
}

@article{greiter2014,
  author = {Greiter, Martin and Schroeter, Darrell F. and Thomale, Ronny},
  title = {{Parent Hamiltonian for the non-Abelian chiral spin liquid}},
  journal = {Phys. Rev. B},
  volume = {89},
  pages = {165125},
  year = {2014},
  doi = {10.1103/PhysRevB.89.165125}
}

@article{SedrakyanGlazmanKamenev2014,
  author  = {Sedrakyan, Tigran A. and Glazman, Leonid I. and Kamenev, Alex},
  title   = {{Absence of {Bose} condensation on lattices with moat bands}},
  journal = {Phys. Rev. B},
  volume  = {89},
  pages   = {201112(R)},
  year    = {2014},
  doi     = {10.1103/PhysRevB.89.201112},
  url     = {https://doi.org/10.1103/PhysRevB.89.201112},
  eprint  = {1303.7272},
  archivePrefix = {arXiv},
  primaryClass = {cond-mat.quant-gas}
}

@article{SedrakyanGlazmanKamenev2015,
  author  = {Sedrakyan, Tigran A. and Glazman, Leonid I. and Kamenev, Alex},
  title   = {{Spontaneous formation of a nonuniform chiral spin liquid in a moat-band lattice}},
  journal = {Phys. Rev. Lett.},
  volume  = {114},
  pages   = {037203},
  year    = {2015},
  doi     = {10.1103/PhysRevLett.114.037203},
  url     = {https://doi.org/10.1103/PhysRevLett.114.037203},
  eprint  = {1409.7359},
  archivePrefix = {arXiv},
  primaryClass = {cond-mat.str-el}
}

@article{MaitiSedrakyan2019,
  author  = {Maiti, Saurabh and Sedrakyan, Tigran A.},
  title   = {{Fermionization of bosons in a flat band}},
  journal = {Phys. Rev. B},
  volume  = {99},
  pages   = {174418},
  year    = {2019},
  doi     = {10.1103/PhysRevB.99.174418},
  url     = {https://doi.org/10.1103/PhysRevB.99.174418},
  eprint  = {1810.00910},
  archivePrefix = {arXiv},
  primaryClass = {cond-mat.str-el}
}

@article{SedrakyanMoessnerKamenev2020,
  author  = {Sedrakyan, Tigran A. and Moessner, Roderich and Kamenev, Alex},
  title   = {{Helical spin liquid in a triangular {XXZ} magnet from {Chern-Simons} theory}},
  journal = {Phys. Rev. B},
  volume  = {102},
  pages   = {024430},
  year    = {2020},
  doi     = {10.1103/PhysRevB.102.024430},
  url     = {https://doi.org/10.1103/PhysRevB.102.024430},
  eprint  = {1911.02932},
  archivePrefix = {arXiv},
  primaryClass = {cond-mat.str-el}
}

@article{WangXieWangSedrakyan2022,
  author  = {Wang, Rui and Xie, Z. Y. and Wang, Baigeng and Sedrakyan, Tigran A.},
  title   = {{Emergent topological orders and phase transitions in lattice {Chern-Simons} theory of quantum magnets}},
  journal = {Phys. Rev. B},
  volume  = {106},
  pages   = {L121117},
  year    = {2022},
  doi     = {10.1103/PhysRevB.106.L121117},
  url     = {https://doi.org/10.1103/PhysRevB.106.L121117},
  eprint  = {2101.04864},
  archivePrefix = {arXiv},
  primaryClass = {cond-mat.str-el}
}

@article{Pachos2004ThreeSpin,
  title   = {{Three-spin interactions in optical lattices and criticality in cluster Hamiltonians}},
  author  = {Pachos, Jiannis K. and Plenio, Martin B.},
  journal = {Phys. Rev. Lett.},
  volume  = {93},
  pages   = {056402},
  year    = {2004},
  doi     = {10.1103/PhysRevLett.93.056402}
}

@article{Chaudhary2019FloquetExchange,
  title   = {{Orbital Floquet Engineering of Exchange Interactions in Magnetic Materials}},
  author  = {Chaudhary, Swati and Hsieh, David and Refael, Gil},
  journal = {Phys. Rev. B},
  volume  = {100},
  pages   = {220403(R)},
  year    = {2019},
  doi     = {10.1103/PhysRevB.100.220403}
}

@article{Kuhr2016QGM,
  title   = {{Quantum-gas microscopes: a new tool for cold-atom quantum simulators}},
  author  = {Kuhr, Stefan},
  journal = {Natl. Sci. Rev.},
  volume  = {3},
  number  = {2},
  pages   = {170--172},
  year    = {2016},
  doi     = {10.1093/nsr/nww023}
}

@article{Yang2021TriangularQGM,
  title   = {{Site-Resolved Imaging of Ultracold Fermions in a Triangular-Lattice Quantum Gas Microscope}},
  author  = {Yang, Jin and Liu, Liyu and Mongkolkiattichai, Jirayu and Schau{\ss}, Peter},
  journal = {PRX Quantum},
  volume  = {2},
  pages   = {020344},
  year    = {2021},
  doi     = {10.1103/PRXQuantum.2.020344}
}

@article{Kuznetsova2023RydbergChiral,
  title   = {{Engineering chiral spin interactions with Rydberg atoms}},
  author  = {Kuznetsova, Elena and Mistakidis, Simeon I. and Rittenhouse, Seth T. and Yelin, Susanne F. and Sadeghpour, H. R.},
  journal = {arXiv preprint},
  eprint  = {2309.08795},
  archivePrefix = {arXiv},
  primaryClass  = {physics.atom-ph},
  year    = {2023}
}

@article{ValenciaTortora2024RydbergChiral,
  title   = {{Rydberg Platform for Nonergodic Chiral Quantum Dynamics}},
  author  = {Valencia-Tortora, Riccardo J. and Pancotti, Nicola and Fleischhauer, Michael and Bernien, Hannes and Marino, Jamir},
  journal = {Phys. Rev. Lett.},
  volume  = {132},
  pages   = {223201},
  year    = {2024},
  doi     = {10.1103/PhysRevLett.132.223201}
}

@article{YangYang1969,
  author = {Yang, C. N. and Yang, C. P.},
  title = {{Thermodynamics of a One-Dimensional System of Bosons with Repulsive Delta-Function Interaction}},
  journal = {J. Math. Phys.},
  volume = {10},
  pages = {1115--1122},
  year = {1969},
  doi = {10.1063/1.1664947}
}

@article{Wen1990ChiralLL,
  author = {Wen, X. G.},
  title = {Chiral {Luttinger} liquid and the edge excitations in the fractional quantum {Hall} states},
  journal = {Phys. Rev. B},
  volume = {41},
  pages = {12838--12844},
  year = {1990},
  doi = {10.1103/PhysRevB.41.12838}
}

@article{NielsenChadha1976,
  author = {Nielsen, H. B. and Chadha, S.},
  title = {On how to count {Goldstone} bosons},
  journal = {Nucl. Phys. B},
  volume = {105},
  pages = {445--453},
  year = {1976},
  doi = {10.1016/0550-3213(76)90025-0}
}

@article{Leutwyler1994,
  author = {Leutwyler, H.},
  title = {Nonrelativistic effective {Lagrangians}},
  journal = {Phys. Rev. D},
  volume = {49},
  pages = {3033--3043},
  year = {1994},
  doi = {10.1103/PhysRevD.49.3033},
  eprint = {hep-ph/9311264},
  archivePrefix = {arXiv}
}

@article{WatanabeMurayama2012,
  author = {Watanabe, Haruki and Murayama, Hitoshi},
  title = {Unified Description of {Nambu}-{Goldstone} Bosons without {Lorentz} Invariance},
  journal = {Phys. Rev. Lett.},
  volume = {108},
  pages = {251602},
  year = {2012},
  doi = {10.1103/PhysRevLett.108.251602},
  eprint = {1203.0609},
  archivePrefix = {arXiv},
  primaryClass = {hep-th}
}

@article{Hidaka2013,
  author = {Hidaka, Yoshimasa},
  title = {Counting Rule for {Nambu}-{Goldstone} Modes in Nonrelativistic Systems},
  journal = {Phys. Rev. Lett.},
  volume = {110},
  pages = {091601},
  year = {2013},
  doi = {10.1103/PhysRevLett.110.091601},
  eprint = {1203.1494},
  archivePrefix = {arXiv},
  primaryClass = {hep-th}
}

@article{ImambekovSchmidtGlazman2012,
  author = {Imambekov, Adilet and Schmidt, Thomas L. and Glazman, Leonid I.},
  title = {One-dimensional quantum liquids: Beyond the {Luttinger} liquid paradigm},
  journal = {Rev. Mod. Phys.},
  volume = {84},
  pages = {1253--1306},
  year = {2012},
  doi = {10.1103/RevModPhys.84.1253},
  eprint = {1110.1374},
  archivePrefix = {arXiv},
  primaryClass = {cond-mat.str-el}
}

@article{MacDonaldGirvinYoshioka1988,
  author  = {MacDonald, A. H. and Girvin, S. M. and Yoshioka, D.},
  title   = {{$t/U$ expansion for the Hubbard model}},
  journal = {Phys. Rev. B},
  volume  = {37},
  pages   = {9753--9756},
  year    = {1988},
  doi     = {10.1103/PhysRevB.37.9753},
  url     = {https://link.aps.org/doi/10.1103/PhysRevB.37.9753}
}

@article{SedrakyanGalitski2010,
  author  = {Sedrakyan, Tigran A. and Galitski, Victor},
  title   = {{Boundary Wess-Zumino-Novikov-Witten model from the pairing Hamiltonian}},
  journal = {Phys. Rev. B},
  volume  = {82},
  pages   = {214502},
  year    = {2010},
  doi     = {10.1103/PhysRevB.82.214502},
  url     = {https://link.aps.org/doi/10.1103/PhysRevB.82.214502},
  eprint  = {1005.0544},
  archivePrefix = {arXiv},
  primaryClass = {cond-mat.str-el}
}

@article{Yuzbashyan2018AOP,
  author  = {Yuzbashyan, Emil A.},
  title   = {{Integrable time-dependent Hamiltonians, solvable Landau--Zener models and Gaudin magnets}},
  journal = {Annals of Physics},
  volume  = {392},
  pages   = {323--339},
  year    = {2018},
  doi     = {10.1016/j.aop.2018.01.017},
  eprint  = {1802.01571},
  archivePrefix = {arXiv},
  primaryClass = {math-ph}
}

@article{BertiniMFT2015,
  author  = {Bertini, Lorenzo and De Sole, Alberto and Gabrielli, Davide and Jona-Lasinio, Giovanni and Landim, Claudio},
  title   = {{Macroscopic fluctuation theory}},
  journal = {Rev. Mod. Phys.},
  volume  = {87},
  pages   = {593--636},
  year    = {2015},
  doi     = {10.1103/RevModPhys.87.593},
  url     = {https://doi.org/10.1103/RevModPhys.87.593}
}

\end{document}